\documentclass[12pt]{article}

\RequirePackage[OT1]{fontenc}
\usepackage{natbib}
\RequirePackage[colorlinks,citecolor=blue,urlcolor=blue]{hyperref}
\usepackage[english]{babel}
\usepackage{amsthm}
\usepackage{amsmath}
\usepackage{amsfonts}
\usepackage{amssymb}
\usepackage{graphicx}
\usepackage{enumerate}
\usepackage{color}
\usepackage{array}
\usepackage{tikz}
\usepackage{comment}
\usepackage{bm}

\definecolor{myorange}{HTML}{E24100}
\definecolor{mygreen}{HTML}{228b22}

\def\frac#1#2{{\textstyle{#1\over#2}}}

\DeclareSymbolFont{AMSb}{U}{msb}{m}{n}
\DeclareMathSymbol{\Natural}{\mathbin}{AMSb}{"4E}
\DeclareMathSymbol{\Integer}{\mathbin}{AMSb}{"5A}
\DeclareMathSymbol{\Real}{\mathbin}{AMSb}{"52}
\DeclareMathSymbol{\Rational}{\mathbin}{AMSb}{"51}
\DeclareMathSymbol{\Imaginary}{\mathbin}{AMSb}{"49}
\DeclareMathSymbol{\Complex}{\mathbin}{AMSb}{"43} 
\DeclareMathSymbol{\Disk}{\mathbin}{AMSb}{"44} 
\def\bi{\begin{itemize}}
\def\ei{\end{itemize}}
\def\bd{\begin{description}}
\def\ed{\end{description}}
\def\ben{\begin{enumerate}}
\def\een{\end{enumerate}}
\def\bv{\begin{verbatim}}
\def\ev{\end{verbatim}}

\def\calC{{\mathcal C}}
\def\calD{{\mathcal D}}

\def\calN{{\mathcal N}}

\def\calS{{\mathcal{S}}}

\def\hat#1{{\widehat{#1}}}

\def\pr{{\rm Pr}}
\def\Pr{\pr}
\def\E{{\rm E}}

\def\2to{{\ {\buildrel 2\over \longrightarrow}\ }}

\def\I1ton{{$I_1,\ldots,I_n$}}
\def\X1ton{{$X_1,\ldots,X_n$}}
\def\Y1ton{{$Y_1,\ldots,Y_n$}}
\def\Z1ton{{$Z_1,\ldots,Z_n$}}
\def\R1ton{{$R_1,\ldots,R_n$}}
\def\e1ton{{$e_1,\ldots,e_n$}}
\def\t1ton{{$t_1,\ldots,t_n$}}
\def\x1ton{{$x_1,\ldots,x_n$}}
\def\y1ton{{$y_1,\ldots,y_n$}}
\def\z1ton{{$z_1,\ldots,z_n$}}

\def\calS{{\mathcal{S}}}

\newcommand\independent{\protect\mathpalette{\protect\independenT}{\perp}}
\def\independenT#1#2{\mathrel{\rlap{$#1#2$}\mkern2mu{#1#2}}}

\begin{document}
\thispagestyle{empty}
\baselineskip=28pt
\vskip 5mm
\begin{center} 
{\Large{\bf Modeling Non-Stationary Temperature Maxima Based on Extremal Dependence Changing with Event Magnitude}}
\end{center}

\baselineskip=12pt
\vskip 5mm

\begin{center}
\large
Peng Zhong$^1$, Rapha\"el Huser$^1$, Thomas Opitz$^2$
\end{center}

\footnotetext[1]{
\baselineskip=10pt Computer, Electrical and Mathematical Sciences and Engineering (CEMSE) Division, King Abdullah University of Science and Technology (KAUST), Thuwal 23955-6900, Saudi Arabia. E-mails: peng.zhong@kaust.edu.sa; raphael.huser@kaust.edu.sa}
\footnotetext[2]{
\baselineskip=10pt BioSP, INRAE, Avignon, 84914, France, E-mail: thomas.opitz@inrae.fr}

\baselineskip=17pt
\vskip 4mm
\centerline{\today}
\vskip 6mm

\begin{center}
{\large{\bf Abstract}}
\end{center}
The modeling of spatio-temporal trends in temperature extremes can help better understand the structure and frequency of heatwaves in a changing climate. Here, we study annual temperature maxima over Southern Europe using a century-spanning dataset observed at 44 monitoring stations.  Extending the spectral representation of max-stable processes, our modeling framework relies on a novel construction of max-infinitely divisible processes, which include covariates to capture spatio-temporal non-stationarities. Our new model keeps a popular max-stable process on the boundary of the parameter space, while flexibly capturing weakening extremal dependence at increasing quantile levels and asymptotic independence. This is achieved by linking the overall magnitude of a spatial event to its spatial correlation range, in such a way that more extreme events become less spatially dependent, thus more localized. Our model reveals salient features of the spatio-temporal variability of European temperature extremes, and it clearly outperforms natural alternative models. Results show that the spatial extent of heatwaves is smaller for more severe events at higher altitudes, and that recent heatwaves are moderately wider. Our probabilistic assessment of the 2019 annual maxima confirms the severity of the 2019 heatwaves both spatially and at individual sites, especially when compared to climatic conditions prevailing in 1950--1975.

\baselineskip=16pt

\par\vfill\noindent

\pagenumbering{arabic}
\baselineskip=24pt

\newpage


\allowdisplaybreaks

\section{Introduction}
\label{sec:introduction}

In the current era of climate change and ecological transitions, environmental risks such as heatwaves or floods are major threats that our society faces more than ever. Available data are becoming increasingly rich and allow us to  develop and implement mathematically sound statistical models to assess the risk associated with spatial environmental extreme events. There is now very broad scientific consensus that global warming is a fact. Climate science research also strongly supports the claim that the nature and magnitude of extreme events undergo a strong evolution due to global change, but that the regional responses may be quite different \citep{IPCC}. We therefore need appropriate statistical models that shed light into the mechanisms leading to extreme episodes in environmental variables, and that can accurately describe their spatio-temporal dependence and variability. As a recent example, the two major heatwaves in June and July of 2019 affected major parts of Europe, which suffered new all-time temperature records established at a large number of weather stations. In this paper, we develop new statistical methodology to study non-stationary extreme temperatures observed over the southern part of Europe from 1918 to 2018, and we then exploit our new model to assess the severity and spatial extent of the 2019 heatwaves. Since the strength of spatial dependence is directly linked to the spatial extent of heatwaves, our analysis can reveal how the spatial scales of 
high temperatures are influenced by local spatial features, temporal trends and the overall magnitude of the event. 

In contrast to traditional statistical models that are appropriate for capturing the ``average" behavior of such phenomena, spatial extreme-value models focus on modeling the joint tails of spatial processes. In this context, max-stable processes have played a central role, being the only possible non-degenerate limits of linearly rescaled pointwise maxima of random processes \citep{Davison.etal:2012,Davison.Huser:2015,Davison.etal:2019}. In practice, max-stable process models are commonly fitted to block maxima, which are often based on annual blocks \citep{Davison.Gholamrezaee:2012}. By contrast to approaches based on high threshold exceedances \citep{Davison.Smith:1990}, the block maximum approach focuses on the long-term behavior of extremes, avoiding the intricate treatment of seasonality and short-term temporal dependence, and does not rely on the (sometimes arbitrary) choice of a threshold discriminating extremes from the bulk, which can be awkward under spatio-temporal non-stationarity \citep{Scarrott.MacDonald:2012}. Popular choices for parametric max-stable models include the Brown--Resnick model \citep{Brown.Resnick:1977,Kabluchko.etal:2009}, or the extremal-$t$ model \citep{Opitz:2013}, which comprises the Schlather model \citep{Schlather:2002} as a special case and the Brown--Resnick model as a limiting case. However, max-stable processes $\{Z(\bm s)\}_{\bm s\in\calS}$ always display a property known as \emph{asymptotic dependence} (except in the trivial case of full independence), which means that the limit $\chi={\lim_{u\to1}\Pr\{Z(\bm s_1)>G_1^{-1}(u) \mid Z(\bm s_2)>G_2^{-1}(u)\}}$, $\bm s_1, \bm s_2\in\calS$, where $G_1$ and $G_2$ denote the cumulative distribution function (CDF) of $Z(\bm s_1)$ and $Z(\bm s_2)$, respectively, exists and is positive ($\chi>0$). This implies that max-stable processes can only capture strong tail dependence, and are inappropriate when maxima---or the original data from which maxima are extracted---are asymptotically independent ($\chi=0$), which corresponds to the situation where the extremal dependence strength eventually weakens and completely vanishes as the quantile level increases (i.e., as $u\to1$). Max-stability is in fact a strong theoretical property that arises asymptotically when considering blocks of increasing size, and which largely restricts the flexibility of extreme-value models (and the class of models considered). As the block size is often chosen to be one year (or less) in real data applications, \emph{imposing} max-stability is often an overly restrictive simplification, which yields an artificially strong extremal dependence structure. This model misspecification is problematic, as it potentially leads to a significant overestimation of joint tail probabilities and thus impacts risk assessment of spatial extreme events. However, while there is a wide body of literature developing peaks-over-threshold models for asymptotic independence or hybrid models bridging the two asymptotic dependence regimes \citep{Wadsworth.Tawn:2012,Opitz:2016,Huser.etal:2017,Huser.Wadsworth:2019,Shooter.etal:2019,Wadsworth.Tawn:2019}, there are only a few papers so far where this problem has been rigorously tackled for block maxima data; see \citet{Bopp.etal:2020} and \citet{Huser.etal:2020}. It is indeed difficult to develop principled sub-asymptotic models for block maxima, which reasonably depart from limiting max-stable processes, while keeping certain properties that reflect the specific type of positive dependence of maxima.

In this paper, we build upon \citet{Bopp.etal:2020} and \citet{Huser.etal:2020}, and develop flexible spatial models that pertain to the wider class of max-infinitely divisible (max-id) processes. Max-id processes naturally extend max-stable processes and relax their rigid dependence structure. While the theory behind max-id processes has been well established for decades \citep{Resnick1987,Gine1990,Dombry2013}, \citet{Padoan2013} was the first to propose a max-id model that has a magnitude-dependent extremal dependence structure. This parametric model stems from taking the limit of block maxima over independent and identically distributed (i.i.d.) Gaussian process ratios, with correlation strength increasing to one as the block size tends to infinity. However, while this model captures asymptotic independence, it is rather inflexible in its ability to capture weakening but strong spatial dependence, and was found to be the worst-performing model fitted in the application of \citet{Huser.etal:2020}. More importantly, this model does not have a max-stable model as a special case, which makes it unable for maxima defined over moderate or large blocks. Alternatively, \citet{Huser.etal:2020} proposed general construction principles for building quite flexible max-id models that remain in the ``neighborhood'' of the extremal-$t$ max-stable process. In particular, they adapted the spectral representation of max-stable processes to construct flexible max-id models that have a smooth transition between asymptotic dependence classes on the boundary of the parameter space. However, the dependence structure of those max-id models remains quite rigid for describing the central part of the distribution of componentwise maxima. Alternatively, \citet{Bopp.etal:2020} recently developed a Bayesian hierarchical max-id model that scales well with large datasets and keeps the \citet{Reich.Shaby:2012} max-stable model as a special case, but whose tail properties are even less flexible than the models proposed by \citet{Huser.etal:2020}. In this paper, we extend the max-id models of \citet{Huser.etal:2020} even further, in order to retain their appealing tail dependence properties and gain significant flexibility in the bulk of the max-id distribution with just one additional parameter. The novel approach that we develop here is to construct max-id processes by taking maxima over random fields 
whose spatial correlation range depends on a random variable representing the overall event magnitude. Furthermore, the max-id models of \citet{Padoan2013}, \citet{Huser.etal:2020} and \citet{Bopp.etal:2020} have stationary and isotropic dependence structures, both in space and time, which is not realistic when modeling environmental data over relatively large areas and long time periods. In this paper, we develop non-stationary max-id models that have a rather parsimonious construction and include spatial and temporal covariates in their dependence structure, 
in order to flexibly capture spatio-temporal variations.
		
The remainder of the paper is organized as follows. In \S\ref{sec:model}, we develop our general modeling framework. More precisely, after giving some background theory about max-id processes, we build a new non-stationary max-id model with a spatial dependence structure that varies according to the three dimensions of (i) space, (ii) time, and (iii) event magnitude. In \S\ref{sec:inference}, we develop our inference approach based on a pairwise likelihood, and demonstrate its good performance with a simulation study. In \S\ref{sec:application}, we further detail our parametric max-id models, and we detail a substantial application of these models to study European temperature extremes, in order to assess the risk of spatial extreme temperatures, such as the 2019 European Heatwaves. Concluding remarks are enclosed in \S\ref{sec:conclusion}.


\section{Modeling based on max-infinitely divisible processes}
\label{sec:model}

\subsection{Marginal modeling of extremes}
\label{sec:margins}
Accurate modeling of marginal distributions and trends in their parameters is paramount to obtaining reliable inferences and predictions on extreme values. We here follow standard limit theory and use the flexible three-parameter generalized extreme value (GEV) distribution for modeling univariate extremes and for incorporating covariate information. Given a sequence of independent and identically distributed random variables  $Y_1,Y_2,\ldots$, the block maximum with block size  $n$ is defined as $Z_n=\max(Y_1,\ldots,Y_n)$. \cite{GEV} showed that if there exist sequences of constants $a_n>0$ and $b_n$ such that the limit variable $\lim_{n\to\infty}{(Z_n-b_n)/a_n}$ has a non-degenerate distribution $G$, then $G$ is from the GEV family, i.e., 
\begin{equation}\label{eq:GEV}
G(z)=\exp\left[-\left\{1+\xi(z-\mu)/\sigma\right\}_+^{-1/\xi}\right],\qquad \xi \neq 0,
\end{equation}
with support $\{z: 1+\xi(z-\mu)/\sigma >0\}$, where $a_+=\max(0,a)$, and the Gumbel distribution $\exp\left[-\exp\left\{-(z-\mu)/\sigma\right\}\right]$, $z\in\Real$, is obtained as $\xi\to0$. Here, $\mu$, $\sigma>0$, and $\xi$ are location, scale and shape parameters, respectively. We distinguish three types of ${\rm GEV}$ distributions depending on the value of $\xi$: Fr\'echet, Gumbel and reversed Weibull, corresponding to $\xi>0$ (heavy-tailed), $\xi\to0$ (light-tailed) and $\xi<0$ (bounded tail), respectively. The $\mathrm{GEV}$ distribution family is the only univariate max-stable distribution, for which there exist constants $a_m>0$ and $b_m$, $m=1,2,\ldots$, such that $G^m(a_mz+b_m)=G(z)$ for all $z$ (using the notation $G^m(z)=\{G(z)\}^m$). In spatial modeling, we can embed covariates in the GEV parameters $(\mu,\sigma,\xi)^T$, or nonlinear covariate effects using spline functions. In our extreme temperature data application detailed in \S\ref{sec:application}, we use cubic regression splines based on spatial coordinates, altitude, and time to flexibly model trends in the marginal distributions. 

\subsection{Max-stable processes and their limitations}
If the multivariate distribution of a $D$-dimensional vector of componentwise maxima of several dependent random variables, linearly renormalized as in \S\ref{sec:margins}, converges to a nondegenerate limiting joint distribution $G$, then it is max-stable in the sense that it satisfies
\begin{equation} \label{max-stable-def}
G^m(\boldsymbol{a}_m\bm z+\boldsymbol{b}_m)= G(\bm z), \quad m=1,2,\ldots
\end{equation}
with normalizing vector sequences $\boldsymbol{a}_m\in(0,\infty)^D$ and $\boldsymbol{b}_m\in\mathbb{R}^D$.  
More generally, a random process $\{Z(\bm s)\}_{\bm s\in\calS}$, defined over the spatial region $\calS\subset\Real^2$, is called max-stable if the property \eqref{max-stable-def} holds for any finite collection of sites $\calD =\{\bm s_1,\ldots,\bm s_D\}\subset\calS$.  	
Given independent copies $\{Y_i(\bm s)\}_{\bm s\in\calS}$, $i=1,2,\ldots$, of a random process $\{Y(\bm s)\}_{\bm s\in\calS}$, we write the spatial process of pointwise maxima as
\begin{equation}\label{eq:limitMS}
    Z_m(\bm s) = \max_{i=1,\dots,m} Y_i(\bm s), \quad \bm s\in\calS.
\end{equation}
If appropriate normalizing sequences $\boldsymbol{a}_m\in(0,\infty)^D$ and $\boldsymbol{b}_m\in\mathbb{R}^D$ exist for all finite configurations of sites $\calD =\{\bm s_1,\ldots,\bm s_D\}\subset\calS$ such that the joint distribution of rescaled pointwise maxima tends to a max-stable limit, then these finite-dimensional distributions define a max-stable process, $\{Z(\bm s)\}_{\bm s\in\calS}$. Therefore, max-stable processes are a natural and popular choice for modeling spatial extremes \citep{Padoan2010,Davison.etal:2012,Davison.etal:2019}. However, as discussed in the introduction in \S\ref{sec:introduction}, max-stable processes are always asymptotically dependent, i.e., unless they are exactly independent, they do \emph{not} allow for the possibility that 
${\lim_{u\to1}\Pr\{Z(\bm s_1)>G_1^{-1}(u) \mid Z(\bm s_2)>G_2^{-1}(u)\}}=0$, 
where $G_1$ and $G_2$ are the CDFs of $Z(\bm s_1)$ and $Z(\bm s_2)$, respectively. Therefore, max-stable processes are not suitable for asymptotically independent data that exhibit non-negligible residual dependence at extreme but finite, sub-asymptotic levels. This limits the applicability of max-stable processes, especially for environmental processes that are often found to exhibit asymptotic independence \citep{Davison.etal:2013,Huser.etal:2017,Huser.Wadsworth:2019,Bacro.al.2019,Bopp.etal:2020}. Using max-stable processes in such a case might lead to substantial overestimation of joint tail probabilities when extrapolating beyond observed levels. Thus, it is sensible to consider the wider class of max-id processes, which contains max-stable processes as a sub-class. Similarly to max-stable processes, max-id processes also arise as certain limits of pointwise maxima \eqref{eq:limitMS}, but where the distribution of the random processes $\{Y_i(\bm s)\}_{\bm s\in\calS}$ in \eqref{eq:limitMS} is allowed to vary with the block size $m$; see \citet{Balkema.Resnick:1977}.

\subsection{Max-infinitely divisible processes}
A random process $\{Z(\bm s)\}_{\bm s\in\calS}$ is called max-infinitely divisible (max-id) if, for any finite collection of sites $\calD=\{\bm s_1,\dots,\bm s_D\}\subset \calS$, the joint distribution $G$ of the random vector $\{Z(\bm s_1),\dots,Z(\bm s_D)\}^T$ is such that $G^t$ defines a valid CDF for any positive real $t>0$. While this is always true in the univariate case ($D=1$) or when $t$ is a positive integer, it may not be true for $D\geq 2$ with non-integer, e.g., fractional, values of $t$. Moreover, in contrast to the max-stable case, $G^t$ does not necessarily stay within the same location-scale family as $G$; this property is only satisfied for the subclass of max-stable distributions; recall \eqref{max-stable-def}.

\citet{Gine1990}, \citet{Resnick1987} and \citet{Haan1993} showed that any max-id process can be constructed by taking pointwise maxima over a Poisson point process (PPP) defined on a suitable function space. Let $\{X_i(\bm s);i=1,2,\dots,N\}_{\bm s\in\calS}$ be the points of a Poisson point process with mean measure $\Lambda$ on the space of continuous functions defined on a compact support $\calS$, denoted by $\calC$, where the measure $\Lambda$ must satisfy certain regularity constraints such as being finite on compact sets; see the above references for details. When $\Lambda(\calC)=\infty$ (such that $N=\infty$ almost surely), we get a max-id process on $\calS$ by setting
\begin{equation}\label{max-id-def}
Z(\bm s) = \sup_{i=1,2,\ldots} X_i(\bm s),\quad \bm s\in\calS. 
\end{equation}
Therefore, max-id processes can be constructed as pointwise maxima over an infinite number of continuous functions  from the space  $\calC$, and the Poisson process weights the functions through its deterministic mean measure $\Lambda$ when sampling from $\calC$.

The mean measure $\Lambda$ is also called the exponent measure of the max-id process, and it determines joint probabilities. Specifically, for a finite number of sites $\calD=\{\bm s_1,\dots,\bm s_D\}\subset \calS$,  the joint distribution $G$ of $\bm Z=\{Z(\bm s_1),\ldots,Z(\bm s_D)\}^T$ is
\begin{equation}\label{eq:joint}
G(\bm z)=\Pr(\bm Z \leq \bm z) = \exp\left\{-\Lambda_\calD([-\bm \infty,\bm z]^C)\right\},\quad \bm z=(z_1,\dots,z_D)^T\in\Real^D,
\end{equation}
where $[-\bm \infty,\bm z] = [-\infty,z_1]\times\cdots\times[-\infty,z_D]\subset\mathbb R^D$, $\Lambda_\calD$ is the restriction of $\Lambda$ to the subspace $\calD\subset\calS$ (i.e., taking measurable sets of $\Real^D$ rather than $\calC$ as input), and $V_\calD(\bm z)=\Lambda_\calD([-\bm \infty,\bm z]^C)$ is called the exponent function. To simplify notation, we henceforth drop the subscript $\calD$ in $V_\calD$ and $\Lambda_\calD$ when no confusion can arise. In the case of max-stable processes, \eqref{max-id-def} can be expressed more specifically through the following spectral construction: 
\begin{equation} \label{max-stable}
    Z(\bm s) = \sup_{i=1,2,\ldots} R_iW_i(\bm s),\quad \bm s \in\calS,
\end{equation}
where $\{R_i;i=1,2,\ldots\}$ are the points of a Poisson point process on the positive half-line $[0,\infty]$ with mean measure $\kappa([r,\infty))=r^{-1}$, $r>0$, and $\{W_i(\bm s)\}_{\bm s\in\calS}$, $i=1,2,\ldots$, are independent copies of a random process $\{W(\bm s)\}_{s\in\calS}$ with $\E[\max\{W(\bm s),0\}]=1$, which are also independent of the points $\{R_i;i=1,2,\ldots\}$; see \citet{Haan1984} and \citet{Schlather:2002}. Let  $\Phi({\rm d} w)$ be the probability distribution associated with the process $\{W(\bm s)\}_{\bm s\in\calS}$ (specified to be Gaussian in our model described below in \S\ref{sec:newmodel}). Hence, the independent random processes $\{X_i(\bm s) = R_iW_i(\bm s);i=1,2,\ldots\}_{\bm s\in\calS}$ are points from a Poisson process with mean measure $\Lambda(A)=\int_{\{rw\in A\}}r^{-2}\text{d}r\Phi({\rm d}w)$, for measurable sets $A\subset\calC$. The exponent function of a max-stable process can be written as $V(\bm z)=\E\left[\max\{W(\bm s_1)/z_1,\ldots,W(\bm s_D)/z_D\}\right]$, $\bm z=(z_1,\ldots,z_D)^T\geq \boldsymbol{0}=(0,\ldots,0)^T$, where $a/0=\infty$ for $a>0$. The max-stable process $\{Z(\bm s)\}_{\bm s\in\calS}$ in \eqref{max-stable} has unit Fr\'echet margins, i.e., $\Pr(Z(\bm s)\leq z)=\exp(-1/z)$, $z>0$.

Using \eqref{eq:joint}, the marginal distribution of $Z$ at a given site $\bm s_0\in\mathcal S$ for general max-id processes is $G_0(z_0)=\exp\left\{-\Lambda_{\bm s_0}(\{z: z > z_0\})\right\}$. To focus on dependence properties, we now assume that the max-id process \eqref{max-id-def} has been standardized using the probability integral transform to have common unit Fr\'echet margins, such that $\Lambda_{\bm s_0}(\{z:z>z_0\})=1/z_0$, $z_0>0$, for all sites $\bm s_0\in\calS$. Then, for any finite collection of sites $\calD=\{\bm s_1,\ldots,\bm s_D\}\subset\mathcal S$, we define the \emph{level-dependent extremal coefficient at (unit Fr\'echet) quantile level $z_0>0$} as 
\begin{equation}\label{extcoef}
\theta_D(z_0) = {\log\{G( \bm z_0)\}\over\log\{G_0(z_0)\}}={\Lambda_\calD([-\bm \infty,\bm z_0]^C)\over\Lambda_{\bm s_0}(\{z:z>z_0\})}=z_0\Lambda_\calD([-\bm \infty,\bm z_0]^C)\in [1,D],
\end{equation}
$\bm z_0 = (z_0,\dots,z_0)^T\in\Real^D$. A similar dependence coefficient was defined by \citet{Padoan2013} and \citet{Huser.etal:2020}. It is easy to see from the definition \eqref{extcoef} that
\begin{equation*}
\Pr(\bm Z\leq \bm z_0) = 
G_0(z_0)^{\theta_D(z_0)}.
\end{equation*}
Therefore, the coefficient $\theta_D(z_0)$ can be interpreted as the effective number of independent variables among $\{Z(\bm s_1),\ldots,Z(\bm s_D)\}^T$ \emph{at quantile level $z_0$}. In the bivariate case, $D=2$, the pair of variables $\bm Z=\{Z(\bm s_1),Z(\bm s_2)\}^T$ turns out to be asymptotically independent if $\lim_{z_0\to\infty}\theta_2(z_0)=2$, and asymptotically dependent otherwise. It can be verified that ${\Pr\{Z(\bm s_2)>z_0\mid Z(\bm s_1)>z_0\}}\sim 2-\theta_2(z_0)$, as $z_0\to\infty$.

With max-stable distributions, the extremal coefficient $\theta_D(z_0)$ is always constant in $z_0$ because the exponent function $V(\bm z)$ is homogeneous of order $-1$, i.e., $V(t \bm z)=t^{-1}V(\bm z)$ for all $t>0$. Thus, max-stable processes cannot capture weakening dependence as events become more extreme. Moreover, they can only capture asymptotic dependence or full independence, but they cannot capture intermediate joint tail decay rates arising with asymptotic independence. The broader class of max-id processes  relaxes such rigid restrictions and yields more flexible models that remain in the ``neighborhood'' of max-stable processes.

\subsection{A new magnitude-dependent max-id model}
\label{sec:newmodel}
For modeling temperature extremes, we build on a max-id construction proposed by \cite{Huser.etal:2020}. It extends the spectral representation of max-stable processes in \eqref{max-stable} and allows capturing asymptotic independence and dependence in a single parametric model. In the max-stable case, the heavy power-law tail of the mean measure $\kappa([r,\infty))=r^{-1}$, $r>0$, of the Poisson process $\{R_i;i=1,2,\ldots\}$, which determines the overall magnitude of the spatial process $\{Z(\bm s)\}_{\bm s\in\calS}$ in \eqref{max-stable}, generates co-occurrences of very large values and leads to asymptotic dependence, while the same level of dependence persists at all quantiles (i.e., $\theta_D(z)$ in \eqref{extcoef} is constant in $z$). 
We can deploy two modifications for the dependence in $Z(\bm s)$ to weaken as the magnitude of extreme events increases. The first modification is to use a lighter-tailed intensity measure $\kappa$ of the Poisson process $\{R_i;i=1,2,\ldots\}$ to attenuate the strong co-occurrence patterns at increasingly high quantiles. The second modification is to relax the independence assumption between the points $\{R_i\}$ and the processes $\{W_i\}$, in such a way to link the spatial dependence range of $W_i$ with the magnitude of $R_i$, which makes the processes $\{W_i\}$ non-identically distributed given $\{R_i\}$. While the first modification was already exploited by \citet{Huser.etal:2020}, the second modification is a new idea. In this paper, we combine both modifications, in order to construct a flexible yet parsimonious max-id model that interpolates between the (asymptotically dependent) extremal-$t$ max-stable model and asymptotic independence with a relatively fast joint probability decay and a flexible form in the bulk.

Following \cite{Huser.etal:2020}, we use a Weibull-tailed mean measure $\kappa$ for $\{R_i\}$ given as
\begin{equation}\label{eq:maxidconstr}
\kappa([r,\infty))=r^{-\beta}\exp\{-\alpha(r^\beta-1)/\beta\},\quad r>0,\, (\alpha,\beta)^T\in (0,\infty)^2.
\end{equation}
We further specify $\{W_i(\bm s)\}_{\bm s\in\calS}$ to be standard Gaussian processes characterized by the correlation function $\rho(\bm s_1,\bm s_2;R_i)$, which may depend on $R_i$. Because $\lim_{\beta\to0}\kappa([r,\infty])=r^{-\alpha}$, this max-id process reduces to the max-stable extremal-$t$ process with $\alpha>0$ degrees of freedom when $\beta\to0$ and $\rho(\bm s_1,\bm s_2;R_i)\equiv \rho(\bm s_1,\bm s_2)$ is independent of $R_i$ \citep{Opitz:2013}. Furthermore, \citet{Huser.etal:2020} showed that when $\{W_i(\bm s)\}_{\bm s\in\calS}$, $i=1,2,\ldots$, are identically distributed standard Gaussian processes with correlation function $\rho(\bm s_1,\bm s_2)$ (independent of $R_i$), the coefficient of tail dependence \citep{Ledford.Tawn:1996}, which characterizes the joint tail decay rate for two sites $\bm s_1,\bm s_2\in\calS$, may be expressed as 
\begin{equation*}
\label{upper_tail_coef}
\eta(\bm s_1,\bm s_2)=\lim_{z\uparrow\infty}{\log\{1-G_1(z)\}\over \log\{1+G(z,z)-2\,G_1(z)\}}=[\{1+\rho(\bm s_1,\bm s_2)\}/2]^{\beta/(\beta+2)},
\end{equation*}
where $G(\cdot,\cdot)$ and $G_1(\cdot)$ represent the bivariate and univariate CDFs of $\{Z(\bm s_1),Z(\bm s_2)\}^T$ and $Z(\bm s_1)$ (or $Z(\bm s_2)$), respectively. Hence, the parameter $\beta$ and the correlation $\rho$ of the Gaussian process $W$ together strongly influence the joint tail decay rate. In particular, as $\beta\to0$ or $\rho(\bm s_1,\bm s_2)\to1$, we get $\eta(\bm s_1,\bm s_2)=1$, which yields asymptotic dependence. In all other cases, we get $\eta(\bm s_1,\bm s_2)<1$, thus asymptotic independence, and we retrieve the tail decay rate of a Gaussian process as $\beta\to\infty$.

\begin{figure}[t!]
        \centering
        \includegraphics[width=\textwidth]{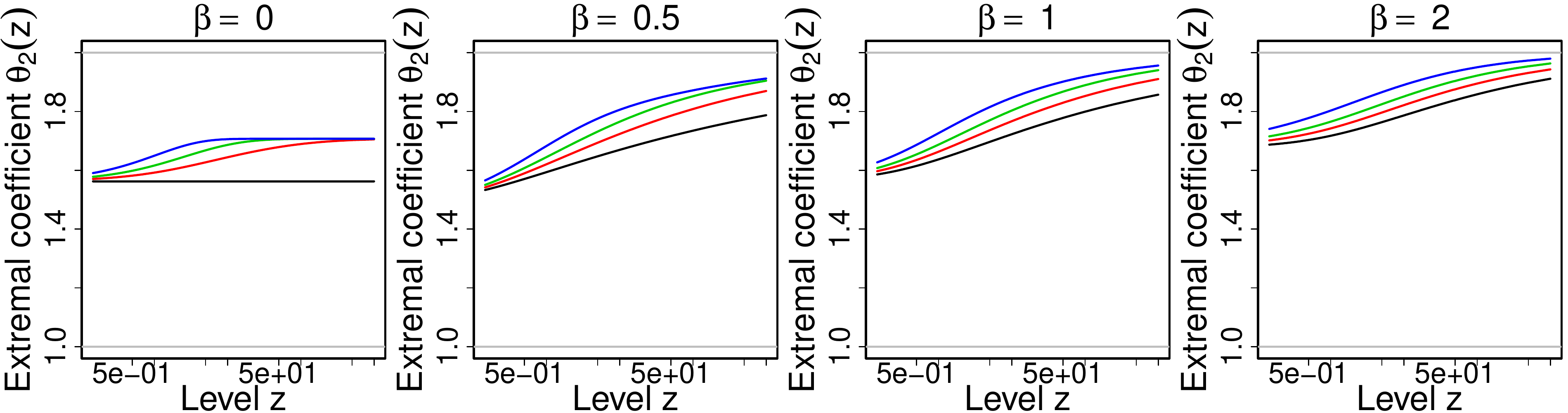}
        \caption{Bivariate level-dependent extremal coefficient $\theta_2(z) = zV(z,z)$ with respect to unit Fr\'echet quantiles $z$, plotted on a logarithmic scale. Our max-id model is defined as in \eqref{max-stable}, where the mean measure $\kappa$ of the Poisson points $\{R_i\}$ is based on \eqref{eq:maxidconstr}, here with $\alpha=1$ and $\beta=0,0.5,1,2$ (left to right), and where the correlation function of the Gaussian processes $W_i$ is here assumed to be $\rho(\bm s_1,\bm s_2;R_i)=\exp\{-\|\bm s_1-\bm s_2\|(1+R_i)^\nu/\lambda\}$ with $\lambda=0.5$ and $\nu = 0$ (black), $\nu = 0.25$ (red), $\nu = 0.5$ (green), $\nu = 1$ (blue). The distance $\|\bm s_1-\bm s_2\|$ is here set to $0.5$. The horizontal grey lines represent the lower and upper bounds of 1 and 2, respectively.}
        \label{fig:ext_coef}
\end{figure}

Here, we extend the model of \citet{Huser.etal:2020} by letting the correlation function of $W_i$ in \eqref{max-stable} depend on $R_i$ such that $\rho(\bm s_1,\bm s_2;R_i)$ decreases as $R_i$ increases. In other words, the spatial dependence strength weakens when the overall event magnitude represented by the points $\{R_i\}$ gets larger. In the stationary and isotropic case, one possibility is to consider the exponential correlation function 
\begin{equation}\label{eq:statcorrelation}
\rho(\bm s_1,\bm s_2;R_i)=\exp\{-\|\bm s_1-\bm s_2\|(1+R_i)^\nu/\lambda\},
\end{equation}
for some baseline range parameter $\lambda>0$, and ``modulation'' parameter $\nu\in\Real$. When $\nu=0$, $\rho(\bm s_1,\bm s_2;R_i)\equiv \rho(\bm s_1,\bm s_2)$ does not depend on $R_i$ (hence retrieving the max-id models of \citet{Huser.etal:2020}), but when $\nu>0$, the spatial range parameter $\lambda(1+R_i)^{-\nu}$ gets smaller (i.e., the dependence strength decreases) as $R_i$ increases (and vice versa when $\nu<0$), with the value of $\nu$ controlling the rate at which the correlation decays with larger points $R_i$. This essentially allows us to get more flexible forms of dependence in the bulk, while keeping appealing tail dependence properties with the \citet{Huser.etal:2020} model as a special case when $\nu=0$. To illustrate the flexibility of this model, Figure~\ref{fig:ext_coef} displays the bivariate level-dependent extremal coefficient $\theta_2(z)$ for various values of $\beta$ and $\nu$. The case $\beta=0$ and $\nu=0$ yields the extremal-$t$ max-stable model, so that $\theta_2(z)$ is constant in $z$. When $\beta=0$ but $\nu>0$, we get asymptotic dependence ($\lim_{z\to\infty}\theta_2(z)<2$) with weakening dependence strength at increasing quantiles. And when $\beta>0$ and $\nu\geq0$, we get asymptotic independence ($\lim_{z\to\infty}\theta_2(z)=2$). Moreover, the extremal coefficient grows with $\nu$ and $\beta$. At any fixed value of $\beta$, the curvature of $\theta_2(z)$ varies significantly for different values of $\nu$, which implies that introducing dependence between $R_i$ and $W_i$ adds considerable flexibility to the model and  improves its ability to appropriately capture the dependence of moderately extreme events. In \S\ref{sec:nonstatcorrel}, we extend this model to the non-stationary, anisotropic case.

By conditioning on the variables $\{R_i\}$, we can prove that the general form of the exponent function of $\{Z(\bm s_1),\ldots,Z(\bm s_D)\}^T$ for our proposed max-id model may be expressed as
\begin{equation}\label{expmeasure}
   V(\bm z) = \Lambda([-\bm \infty,\bm z]^C) = \int_0^\infty \{1-\Phi(\bm z/r;r)\}\,\kappa({\rm d}r)
\end{equation}
where $\Phi(\cdot;r)$ is the joint distribution of the Gaussian vector ${\{W_i(\bm s_1),\ldots,W_i(\bm s_D)\}^T\mid \{R_i=r\}}$ with correlation $\rho(\bm s_1,\bm s_2; r)$. Partial and full derivatives of the exponent function, which are required for likelihood-based inference, can be obtained by differentiating \eqref{expmeasure} with respect to the components of $\bm z$ under the integral sign. Standard formulas, similar to those derived in \citet{Huser.etal:2020}, can be easily obtained, although they are expressed in terms of uni-dimensional integrals that have to be numerically approximated in practice. 

\subsection{Non-stationary dependence structure}\label{sec:nonstatcorrel}
Over large study areas or long periods of time, the strength of extremal dependence, and therefore the spatial extent of clusters of extreme values, may vary. We here extend the exponential correlation model presented in \S\ref{sec:newmodel} to the non-stationary context, and we show how spatio-temporal covariates may be naturally incorporated. We now index the correlation function of the process $W_i$ in \eqref{max-stable} by time $t$ as $\rho_t(\bm s_1,\bm s_2;R_i)$ to emphasize that it may vary over time. Building upon \citet{Paciorek.Schervish:2006} and \citet{Huser2016}, such a non-stationary correlation function on $\Real^2$ may be obtained as follows:
\begin{equation}\label{eq:nonstatcorrelation}
\rho_t(\bm s_1,\bm s_2;R_i) = |\Omega_{\bm s_1,t}(R_i)|^{1/4}|\Omega_{\bm s_2,t}(R_i)|^{1/4}\left|{\Omega_{\bm s_1,t}(R_i)+\Omega_{\bm s_2,t}(R_i)\over 2}\right|^{-1/2}C\{Q_{\bm s_1;\bm s_2,t}^{1/2}(R_i)\},\\
\end{equation}
where $\Omega_{\bm s,t}(R_i)$ is a $2$-by-$2$ covariance matrix that may depend on spatial location $\bm s$, time $t$, and the Poisson points $\{R_i\}$, $C(h)$ is a stationary isotropic correlation function with unit range, e.g., $C(h)=\exp(-h)$, $h\geq0$, and $Q_{\bm s_1;\bm s_2,t}(R_i)$ is the quadratic form
\begin{equation*}
 Q_{\bm s_1;\bm s_2,t}(R_i) = (\bm s_1-\bm s_2)^T\left\{{\Omega_{\bm s_1,t}(R_i)+\Omega_{\bm s_2,t}(R_i)\over2}\right\}^{-1}(\bm s_1-\bm s_2).
\end{equation*}
Covariates, such as time and altitude as used in our temperature data application in \S\ref{sec:application}, can be linked to the matrix $\Omega_{\bm s,t}(R_i)$. As explained in \S\ref{sec:newmodel}, we also allow the variables $\{R_i\}$ to directly influence the range of spatial dependence, which is in contrast with \citet{Huser2016}. More precisely, we propose the following general model for the covariance matrix $\Omega_{\bm s,t}(R_i)$:
\begin{equation}\label{eq:Omega}
\Omega_{\bm s,t}(R_i) = \lambda_{\bm s,t}^2(1 + R_i)^{-2\nu}A(\theta),\quad A(\theta)=\begin{bmatrix} \cos(\theta) & -\sin(\theta)  \\ \sin(\theta) & \cos(\theta) \end{bmatrix}\begin{bmatrix} 1 & 0  \\ 0 & a \end{bmatrix}\begin{bmatrix} \cos(\theta) & -\sin(\theta)  \\ \sin(\theta) & \cos(\theta) \end{bmatrix}^T,
\end{equation}
where $\lambda_{\bm s,t}>0$ is a baseline range parameter that may vary over space and time, and $\nu\in\Real$ as in \S\ref{sec:newmodel}, $a>0$ is a geometric anisotropy scaling that controls the ratio of principal axes of elliptical correlation contours, and $\theta\in[0,\pi/2]$ is a rotation angle of these elliptical contours. The value $a=1$ corresponds to the isotropic case, with $A(\theta)$ reducing to the $2$-by-$2$ identity matrix $I_{2\times 2}$, such that $\Omega_{\bm s,t}(R_i) = \lambda_{\bm s,t}^2(1 + R_i)^{-2\nu}I_{2\times 2}$ and thus \eqref{eq:nonstatcorrelation} corresponds to the correlation \eqref{eq:statcorrelation} if $\lambda_{\bm s,t}\equiv\lambda>0$. To capture spatio-temporal variations in the dependence structure, covariates may be included in $\lambda_{\bm s,t}$. For example, in our real data application in \S\ref{sec:application}, we specify $\lambda_{\bm s,t} = \exp(\lambda_0+\lambda_1\times \text{alt}_{\bm s}+\lambda_2\times t)$, where $\lambda_0,\lambda_1,\lambda_2\in\Real$ are range parameters corresponding to the intercept, the effect of altitude, and the effect of time, respectively, on the spatial dependence range. More precisely, while $\lambda_0$ measures the overall strength of spatial dependence, the parameters $\lambda_1$ and $\lambda_2$ determine whether the dependence structure, and thus the spatial extent of extreme events, changes according to altitude and time, respectively. Several sub-models of \eqref{eq:Omega} described in Table~\ref{tab:sub-models} may be of interest. In \S\ref{sec:application}, we specifically focus on the non-stationary, but locally isotropic case ($a=1$) with $R_i\not\independent W_i$ (i.e., $R_i$ and $W_i$ dependent of each other with $\nu\neq0$), which already yields a rich class of models capturing complex dependence patterns.

\begin{table}[t!]
\centering
\caption{Interesting special cases of Model \eqref{eq:Omega}, categorized into stationary/non-stationary and (locally) isotropic/anisotropic models with $R_i$ independent/dependent of $W_i$ in \eqref{max-stable}.}
\vspace{5pt}
\begin{tabular}{ccc|c}
Stationarity & Anisotropy & Value of $\nu$ & Model type \\ \hline
$\lambda_{\bm s,t}\equiv \lambda$ & $a=1$ & $\nu=0$ & Stationary, isotropic, $R_i\independent W_i$\\
$\lambda_{\bm s,t}\equiv \lambda$ & $a=1$ & $\nu\neq0$ & Stationary, isotropic, $R_i\not\independent W_i$\\
$\lambda_{\bm s,t}\equiv \lambda$ & $a\neq1$ & $\nu=0$ & Stationary, anisotropic, $R_i\independent W_i$\\
$\lambda_{\bm s,t}\equiv \lambda$ & $a\neq1$ & $\nu\neq0$ & Stationary, anisotropic, $R_i\not\independent W_i$ \\
$\lambda_{\bm s,t}\not\equiv \lambda$ & $a=1$ & $\nu=0$ & Non-stationary, locally isotropic, $R_i\independent W_i$\\
$\lambda_{\bm s,t}\not\equiv \lambda$ & $a=1$ & $\nu\neq0$ & Non-stationary, locally isotropic, $R_i\not\independent W_i$\\
$\lambda_{\bm s,t}\not\equiv \lambda$ & $a\neq 1$ & $\nu=0$ & Non-stationary, locally anisotropic, $R_i\independent W_i$\\
$\lambda_{\bm s,t}\not\equiv \lambda$ & $a\neq 1$ & $\nu\neq 0$ & Non-stationary, locally anisotropic, $R_i\not\independent W_i$
\end{tabular}  
\label{tab:sub-models}
\end{table}


\section{Inference using the pairwise likelihood approach}
\label{sec:inference}

\subsection{Two-step modeling of marginal distributions and dependence}\label{sec:twostepinference}
We use a two-step estimation method that is known as ``inference functions for margins" in the literature, and for which consistency and asymptotic normality have been established under mild conditions \citep{Joe1996,Joe2005,Joe2015}.  In the first step, we model only marginal GEV distributions \eqref{eq:GEV} with covariates or semi-parametric spline functions using an independence composite likelihood \citep{Varin.etal:2011}, which is built under the working assumption that the data are spatially independent (given the covariates). We then use the fitted marginal GEV distribution functions to transform the observed data to pseudo-uniform ${\rm Unif}(0,1)$ scores through the probability integral transform. In the second step, we fit the dependence structure (i.e., the copula) of the max-id dependence model to the transformed data using a pairwise likelihood approach, treating the margins as exactly ${\rm Unif}(0,1)$.

\subsection{Pairwise likelihood approach} \label{sec:PL}

Pairwise likelihood has become the standard inference technique for max-stable models owing to the computational intractability of full likelihood expressions in high dimensions \citep{Padoan2010,Padoan2013,Huser2013,Huser.etal:2016,Castruccio.etal:2016,Huser.etal:2019}. The pairwise likelihood approach offers tools akin to classical likelihood inference, is much faster than a full likelihood approach, and usually retains high efficiency. We here adapt this approach to our max-id models. 
Let $\{\bm z_k=(z_{k1},\dots,z_{kD})^T\}_{k=1}^n$ be $n$ independent replicates of the max-id process $Z(\boldsymbol{s})$ with parameter vector $\bm \psi\in\Psi\subset\mathbb{R}^p$ observed at locations $\calD=\{\bm s_1,\dots,\bm s_D\}\subset \calS$. From \eqref{eq:joint}, the full likelihood is 
\begin{equation}\label{FL}
L(\bm\psi;\bm z_1,\dots,\bm z_n) = \prod_{k=1}^n\bigg[\exp\{-V(\bm z_k)\}\sum_{\pi\in\mathcal{P}_D}\prod_{l=1}^{|\pi|}\left\{-V_{\tau_l}(\bm z_k)\right\}\bigg],
\end{equation}
where $\mathcal{P}_D$ is the collection of all partitions $\pi=\{\tau_1,\ldots,\tau_{|\pi|}\}$ sets of $\{1,\dots,D\}$, and $V_{\tau_l}(\bm z_k)$ denotes the partial derivatives of the exponent function $V(\bm z_k)$ with respect to the variables $\{z_{kj}\}_{j\in\tau_l}$, $\tau_l\in\pi$; see, e.g., \citet{Huser.etal:2019}. The number of terms in the sum in \eqref{FL} grows super-exponentially with $D$. The pairwise likelihood approach eases the computational burden by maximizing the pairwise likelihood function $PL(\bm\psi;\bm z_1,\dots, \bm z_n)$ defined as 
\begin{equation}\label{PL}
  \prod_{1\leq j_1 < j_2\leq D}\bigg[\prod_{k=1}^n \exp\{-V(z_{kj_1},z_{kj_2})\}\{V_1(z_{kj_1},z_{kj_2})V_2(z_{kj_1},z_{kj_2})-V_{12}(z_{kj_1},z_{kj_2})\}\bigg]^{\omega_{j_1,j_2}},
\end{equation}
where $\omega_{j_1,j_2}\geq0$ are non-negative weights attributed to the pairs $\{j_1,j_2\}$. Here, we fit the marginal distribution first and compute the pseudo-uniform scores $u_{kj} = \hat{G}_{kj}(z_{kj})$, where $\hat{G}_{kj}$ is the fitted marginal distribution for the $k$-th time point and the $j$-th site $\bm s_j$. Let $\hat g_{kj}$ be the corresponding fitted marginal density. The pairwise likelihood function $PL(\bm \psi;\bm u_1,\dots, \bm u_n)$ based on pseudo-uniform scores $\{\bm u_k=(u_{k1},\dots,u_{kD})^T\}_{k=1}^n$ may thus be written as 
\begin{align}\label{PL_new}
PL(\bm \psi;&\bm u_1,\dots, \bm u_n) =\prod_{1\leq j_1 < j_2\leq D}\bigg(\prod_{k=1}^n \exp\left[-V\{\hat{G}_{kj_1}^{-1}(u_{kj_1}),\hat{G}_{kj_2}^{-1}(u_{kj_2})\}\right]\times\\ 
&\times \left[V_1\{\hat{G}_{kj_1}^{-1}(u_{kj_1}),\hat{G}_{kj_2}^{-1}(u_{kj_2})\}\, V_2\{\hat{G}_{kj_1}^{-1}(u_{kj_1}),\hat{G}_{kj_2}^{-1}(z_{kj_2})\}-V_{12}\{\hat{G}_{kj_1}^{-1}(u_{kj_1}),\hat{G}_{kj_2}^{-1}(u_{kj_2})\}\right]\times\nonumber\\
&\times\left[\hat{g}_{kj_1}\{\hat{G}_{kj_1}^{-1}(u_{kj_1})\}\hat{g}_{kj_2}\{\hat{G}_{kj_2}^{-1}(u_{kj_2})\}\right]^{-1}\bigg)^{\omega_{j_1,j_2}}.\nonumber
\end{align}

Different approaches can be used to select the pairwise likelihood weights $\omega_{j_1,j_2}$, e.g., using binary weights $\omega_{j_1,j_2}\in\{0,1\}$ fixed according to the distance between sites, in order to improve both the computational and statistical efficiency \citep[see, e.g.,][]{Castruccio.etal:2016}. In our simulation study in \S\ref{bs}, we choose $\omega_{j_1,j_2}=I(\|\bm s_{j_1}-\bm s_{j_2}\|\leq \delta)$ for some cutoff distance $\delta>0$ for computational reasons, where $I(\cdot)$ is the indicator function, whereas in our application in \S\ref{sec:application} we use the pragmatic approach of setting $\omega_{j_1,j_2}=1$ for all pairs $\{j_1,j_2\}$. 

It is known that under mild regularity conditions, the pairwise likelihood estimator maximizing \eqref{PL} with known margins is strongly consistent and asymptotically normal with the Godambe variance-covariance matrix, which could in principle be used to assess the variability of the estimator; see, e.g., \citet{Varin.etal:2011} and \citet{Padoan2010} for the max-stable case. A similar asymptotic behavior holds for the estimator $\bm{\hat\psi}$ based on the two-step estimator \eqref{PL_new} (with unknown margins), though the asymptotic variance is generally slightly larger due to the uncertainty in estimating marginal distributions; see, e.g., \citet{Genest.etal:1995} who treat the case where margins are estimated non-parametrically, \citet{Joe1996} for the parametric case, and \citet{Huser.Davison:2014} and \citet{Huser.etal:2016} who compare various parametric estimation schemes for extremes, including one-step and two-step pairwise likelihood estimators. However, since the computation of the asymptotic variance is intricate and may be biased when the data contain many missing values, we here rely on a parametric bootstrap procedure to assess the estimation uncertainty: we repeatedly sample maxima data at the data locations from the fitted max-id model (with the same sample size and with the same number of missing values inserted as in the original dataset), and we then re-estimate parameters using the same pairwise likelihood. Using $300$ bootstrap samples, we can then approximate the distribution and variability of estimated parameters.

\subsection{Simulation experiments}\label{bs}
We conducted a simulation study, in order to assess the performance of the pairwise likelihood estimator $\bm{\hat\psi}$ under a non-stationary setting that resembles our real data application in \S\ref{sec:application}. 

\begin{figure}[t!]
        \centering 
        \includegraphics[width=0.85\textwidth]{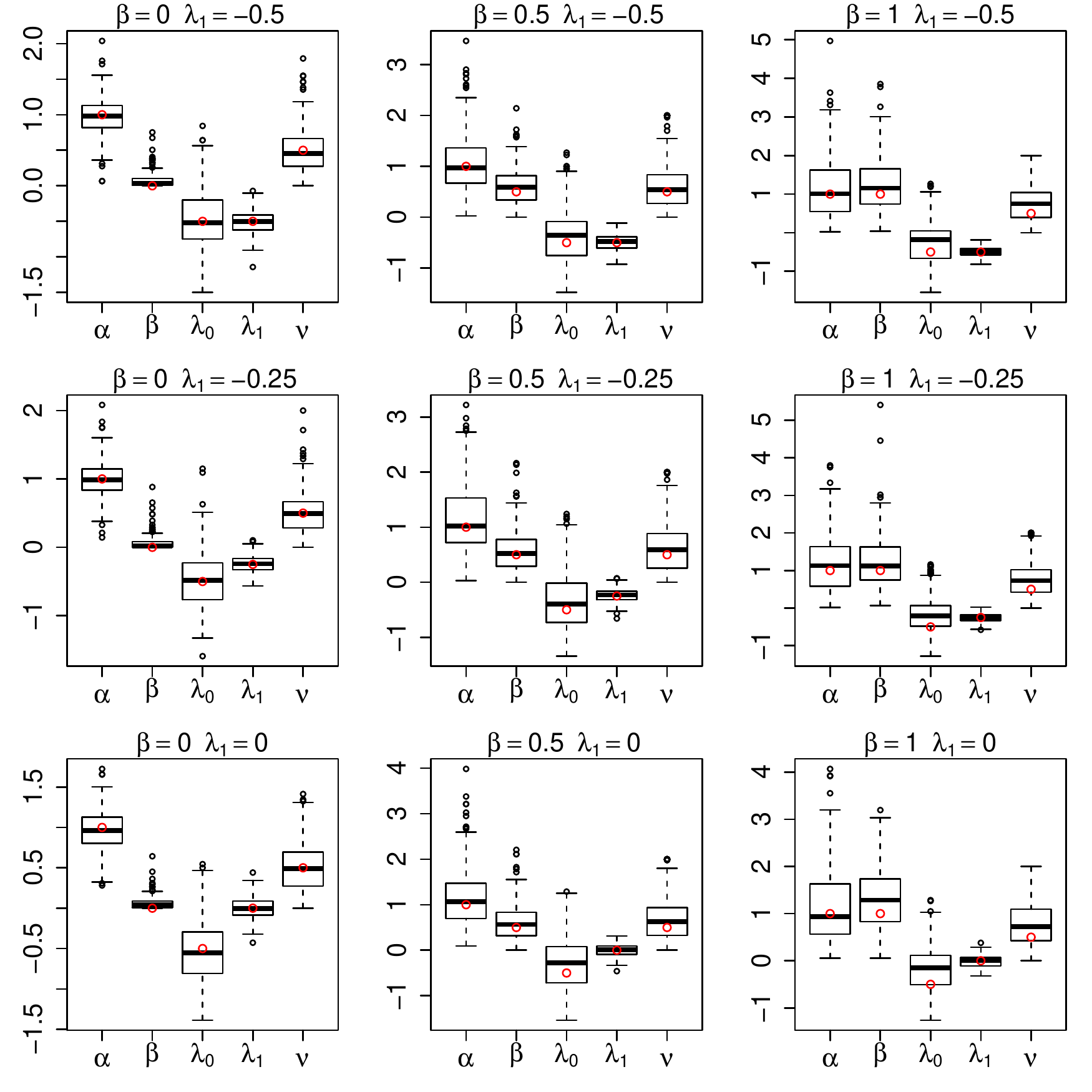}
        \caption{Boxplots of estimated parameters for the simulation study described in \S\ref{bs}. Each panel corresponds to a different simulation scenario with $\lambda_1=-0.5,-0.25,0$ (top to bottom) and $\beta=0,0.5,1$ (left to right) (see details in the text), and shows boxplots for each of the 5 parameters based on 200 experiments. Red dots indicate the true values.}
        \label{fig:fit_simu}
\end{figure}

We simulated data from our proposed max-id model, built from \eqref{max-stable} using the mean measure \eqref{eq:maxidconstr} and the non-stationary correlation function \eqref{eq:nonstatcorrelation} combined with \eqref{eq:Omega}, on the domain $\mathcal S=(0,1)^2$. Here, we focused on the non-stationary, but locally isotropic case (recall Table~\ref{tab:sub-models}), where $\Omega_{\bm s,t} = \lambda_{\bm s,t}^2(1+R_i)^{-2\nu}I_{2\times2}$, with $I_{2\times 2}$ the identity matrix, and we only considered non-stationarity in space such that $\lambda_{\bm s,t}\equiv \lambda_{\bm s}$ only varies with spatial location $\bm s\in\calS$. To mimic the effect of a ``mountain range'', we used the covariate defined as $x_{\bm s}=2\phi(s_x;0.5,0.25)-1$, where $\bm s=(s_x,s_y)^T$ and $\phi(\cdot;\mu,\sigma)$ denotes the Gaussian $\calN(\mu,\sigma^2)$ density, and then defined $\lambda_{\bm s,t}\equiv\lambda_{\bm s} = \exp(\lambda_0+\lambda_1\,x_{\bm s})$, with $\lambda_0 = -0.5$ and $\lambda_1\in\{-0.5,-0.25,0\}$. Negative values of $\lambda_1$ correspond to weaker dependence at higher ``altitudes'', here represented by the covariate $x_{\bm s}$. For the mean measure of the Poisson points $\{R_i\}$ defined in \eqref{eq:maxidconstr}, we chose $\alpha=1$ and $\beta\in\{0,0.5,1\}$ (from asymptotic dependence with $\beta=0$ to asymptotic independence with $\beta>0$), while we selected $\nu=0.25$ to control the interaction between the points $\{R_i\}$ and the processes $\{W_i\}$ in \eqref{max-stable}. Overall, this yields 9 simulation scenarios (3 values of $\lambda_1$ times 3 values of $\beta$), and we then jointly estimated the 5 dependence parameters, namely $\bm\psi=(\alpha,\beta,\lambda_0,\lambda_1,\nu)^T\in(0,\infty)^2\times\Real^3$, treating margins as known here for simplicity. For each case and each dataset, we generated $50$ independent replicates of the process at $49$ sites on $\mathcal S$, roughly located on a $7\times7$ grid with some small additional random perturbations. The dependence parameters were jointly estimated using the pairwise likelihood approach described in \S\ref{sec:PL} with the pairwise likelihood weights $\omega_{j_1,j_2}$ set to be $0$ when the distance between the sites $\bm s_{j_1}$ and $\bm s_{j_2}$ exceeds the cutoff distance $\delta=0.375$ and $\omega_{j_1,j_2}=1$ otherwise, in order to ease the computation burden. We repeated the above steps $200$ times to assess the variability and bias of estimated parameters. Figure~\ref{fig:fit_simu} reports the results and shows boxplots of estimated parameters, with one display for each scenario (i.e., for each pair of values $\{\lambda_1,\beta\}$). 
The parameters are well estimated overall without any strong biases or notable outliers, which suggests that they are well identifiable. The true values (red dots) are close to the medians and always within the interquartile range in all scenarios. While the estimated parameters for $\alpha$ and $\beta$ seem quite variable, especially when $\beta$ is large, the covariate effect $\lambda_1$ always has a fairly moderate variability. Finally, we can notice that the estimated values of $\nu$ are always positive even if the domain of definition for this parameter is fixed to the whole real line in our implementation. This shows that it is easy to identify that the dependence strength is weakening (rather than strengthening) as the severity of extreme events increases.


\section{Application to European temperature extremes}
\label{sec:application}

\subsection{Dataset}\label{sec:dataset}
\begin{figure}[t!]
    \centering
    \includegraphics[width=0.8\textwidth]{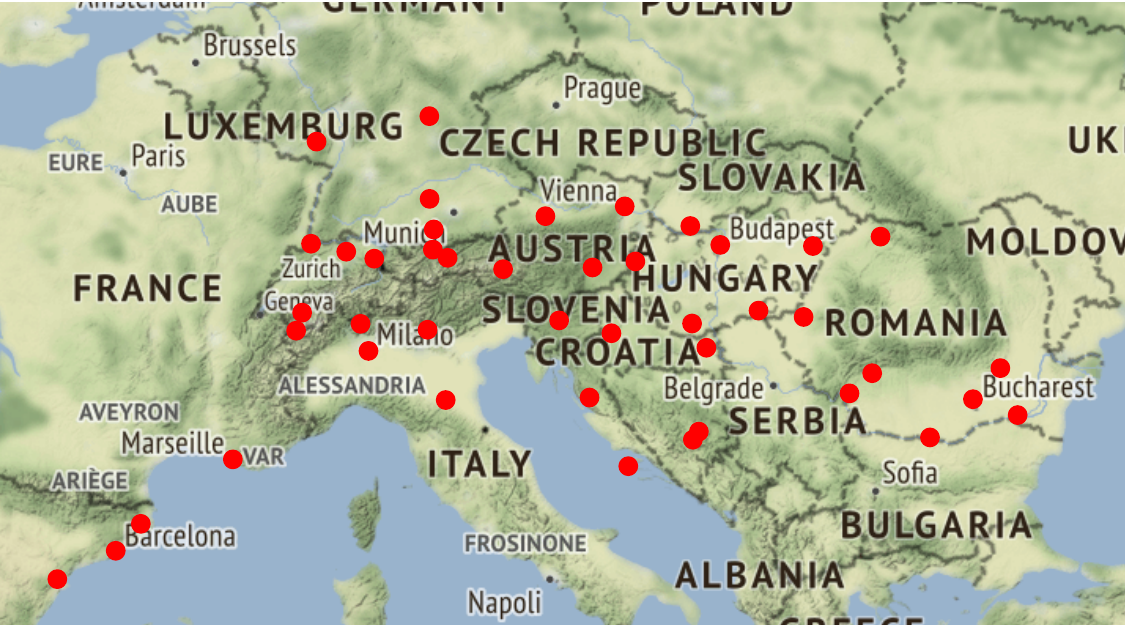}
    \caption{Study region, with the 44 monitoring stations (red dots) distributed over Europe.}
    \label{fig:map}
\end{figure}

In our application, we use the dataset of \citet{dataset} and extract annual maximum temperatures for the period 1918--2018 at $D=44$ monitoring stations in Europe covering a belt between latitudes $40^\circ$ and $50^\circ$ from Western to Eastern Europe, with the Alp mountain range in its central part; see Figure~\ref{fig:map}. This dataset contains $22$ stations with complete records (i.e., without missing values), while missing values account for about $14.7\%$ of observation points overall. Available covariates are the geographical information about the monitoring stations including longitude, latitude and altitude (in km).  We used the great-circle metric (geodesic distance) to compute the distance between sites, which ranges between $22.6$km and $2227.42$km for the different pairs of stations. In the following, we divide distances by $1000$. As our dataset is available for a long period of time comprising several major events (such as the 2003 European heatwave), and at a decent number of monitoring sites with a reasonable spatial coverage over Southern Europe, we here use it to assess how the heatwave risk varies over both space and time, and whether the spatial extent of such extreme phenomena has become wider due to climate change. Understanding whether the severity and spatial extent of extreme temperature events has changed---and where---is indeed key in practice for designing regional mitigation measures of future extreme events. In \S\ref{sec:2019heatwave}, we also assess whether the extreme temperatures observed during the 2019 heatwaves (the first one occurred around end of June, the second one around end of July) over major parts of Europe could have been anticipated from past extreme events.
 
\subsection{Spatial and temporal trends in marginal distributions}\label{sec:spatialtrends}
\begin{figure}[t!]
    \centering
   \includegraphics[width=\textwidth]{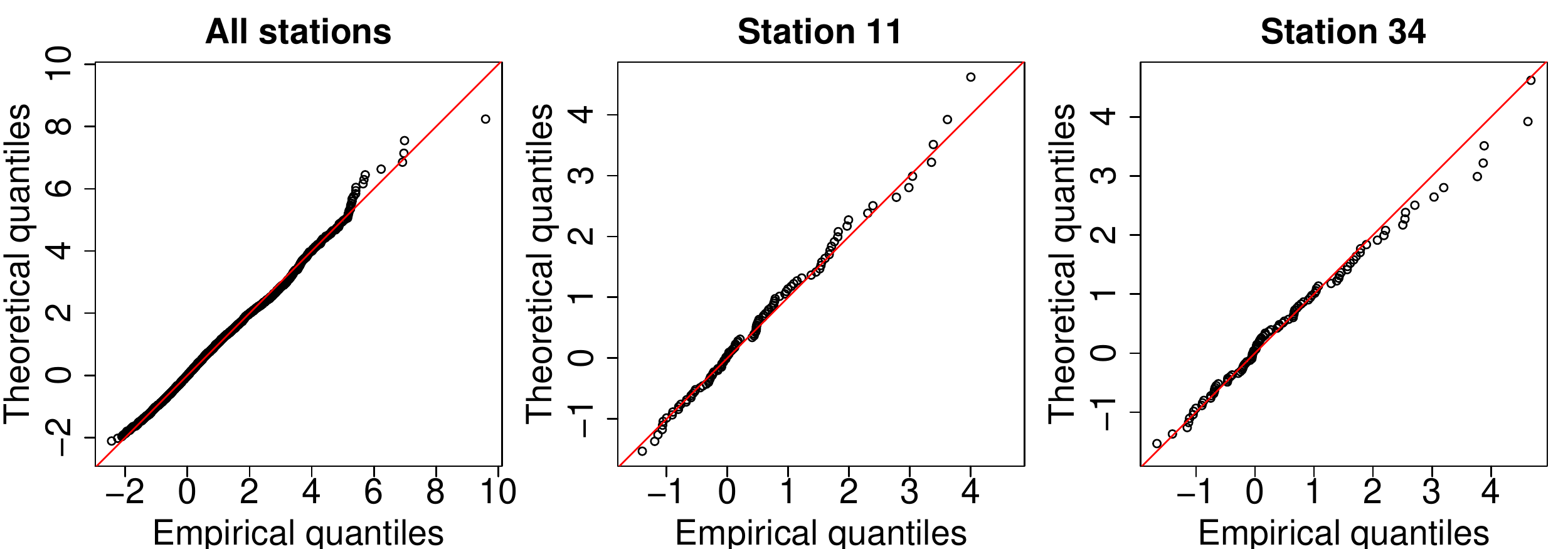}
    \caption{QQ-plots of maxima transformed to the standard Gumbel scale based on the fitted marginal model.  Left: all stations pooled together. Middle and right: stations $\bm s_{11}$ and $\bm s_{34}$. }
    \label{fig:qq_plot}
\end{figure}

The first step of our statistical analysis is to adequately model the non-stationary marginal distributions of maxima. We assume the annual temperature maxima follow a generalized extreme-value (GEV) distribution \eqref{eq:GEV} with parameters potentially depending on longitude, latitude, altitude, as well as time. To reduce the uncertainty in estimated marginal parameters, we pool the data together in single generalized additive model with penalized cubic regression splines to accurately describe the time trend and spatial variation of GEV parameters, and then estimate parameter surfaces by maximizing an independence composite likelihood \citep{Varin.etal:2011}. As explained in \S\ref{sec:twostepinference}, this approach provides valid inference for marginal parameters. Specifically, let $\{Z(\bm s,t)\}_{\bm s\in\calS,t\in[0,1]}$ denote the spatio-temporal process of annual maxima (defined over the space-time domain $\calS\times[0,1]$), and let $Z_{kj}=Z\{\bm s_j,k/(n+1)\}$ be the annual maximum for the $k$-th year at the $j$-th station ($k=1,\ldots,n$, $j=1,\ldots,D$). We assume that $Z(\bm s,t)$ has a marginal GEV distribution with location parameter $\mu_{\bm s,t}$, and constant scale and shape parameters, $\sigma>0$ and $\xi$, respectively. Since the location parameter $\mu_{\bm s,t}$ determines the overall magnitude of extreme values, we link it with the covariates $\text{lon}_{\bm s}$, $\text{lat}_{\bm s}$, $\text{alt}_{\bm s}$ representing longitude, latitude and altitude, respectively, and (rescaled) time $t\in[0,1]$. We also tried to let $\sigma$ vary over space and time, but it did not improve the model significantly. After some experiments, we thus formulated the marginal model as 
\begin{equation}\label{margins}
 Z(\bm s,t)\sim \text{GEV}(\mu_{\bm s,t},\sigma,\xi),\quad \mu_{\bm s,t} = \text{ti}(\text{lon}_{\bm s},\text{lat}_{\bm s},\text{alt}_{\bm s}) + \text{ti}(t), \quad \bm s\in\calS,t\in[0,1],
\end{equation}
where ``${\rm ti}$'' refers to the tensor product of penalized cubic regression splines. This marginal model was chosen to provide a good balance between flexibility (for a good model fit) and parsimony (for robustness and to avoid overfitting), and diagnostics described below suggest that the margins are appropriately modeled over the study region. For each observation, \eqref{margins} yields $Z_{kj}\sim \text{GEV}(\mu_{kj},\sigma,\xi)$, where $\mu_{kj}=\mu_{\bm s_j,k/(n+1)}$, and we fit the marginal model jointly combining all observations by pretending that the $Z_{kj}$'s are independent. Because the geographical location is jointly determined by $\text{lon}_{\bm s}$, $\text{lat}_{\bm s}$ and $\text{alt}_{\bm s}$, they are put together in \eqref{margins} to account for interaction effects, while we keep the time $t$ separate to avoid an overly complex model with too many spline coefficients to be estimated. Here we take 4 spline knots for each dimension, which is rich enough to provide good marginal fits as demonstrated below. Therefore, $\text{ti}(\text{lon}_{\bm s},\text{lat}_{\bm s},\text{alt}_{\bm s})$ has $4^3=64$ spline knots in total. The estimated scale and shape parameters are $\hat\sigma=17.7$ with $95\%$ confidence interval $(17.2,18.1)$ and $\hat\xi=-0.19$ with $95\%$ confidence interval $(-0.22,-0.18)$, respectively. Because $\hat\xi$ is negative, the temperature distribution is estimated to have a finite upper endpoint, which is meaningful in view of the results obtained in similar studies about extreme temperatures \citep[see, e.g.,][]{Davison.Gholamrezaee:2012,Huser2016}. The estimated endpoint $\hat\mu_{\bm s,t}-\hat\sigma/\hat\xi$ varies with the covariates (longitude, latitude, altitude, time) according to the estimated location surface $\hat\mu_{\bm s,t}$. To check the marginal goodness-of-fit, we then transform the maxima $\hat Z_{kj}$ to the standard Gumbel scale as $\hat\xi^{-1}\log\{1+\hat\xi(Z_{kj}-\hat\mu_{kj})/\hat\sigma\}$ by plugging in estimated parameters $\hat\mu_{\bm s,t},\hat\sigma,\hat\xi$, and we produce marginal quantile-quantile (QQ)-plots based on theoretical and empirical standard Gumbel quantiles for each station, and by pooling all stations together. Figure~\ref{fig:qq_plot} displays QQ-plots for the pooled dataset and two randomly selected stations ($\bm s_{11}$ and $\bm s_{34}$). Overall, the marginal goodness-of-fit looks satisfactory, with the dots well aligned along the main diagonal for the vast majority of stations. 
To further examine the quality of the marginal fit, we perform a two-sided Kolmogorov-Smirnov test for the data from each station and for the pooled dataset, by comparing the empirical distributions and the fitted distributions. All the Kolmogorov-Smirnov tests fail to reject the null hypothesis (i.e., equality of distributions) with large p-values. Most of the p-values are greater than $0.9$, and the smallest is 0.28, which confirms good marginal fits overall at all monitoring stations.

\begin{figure}[t!]
\centering
\includegraphics[width=\textwidth]{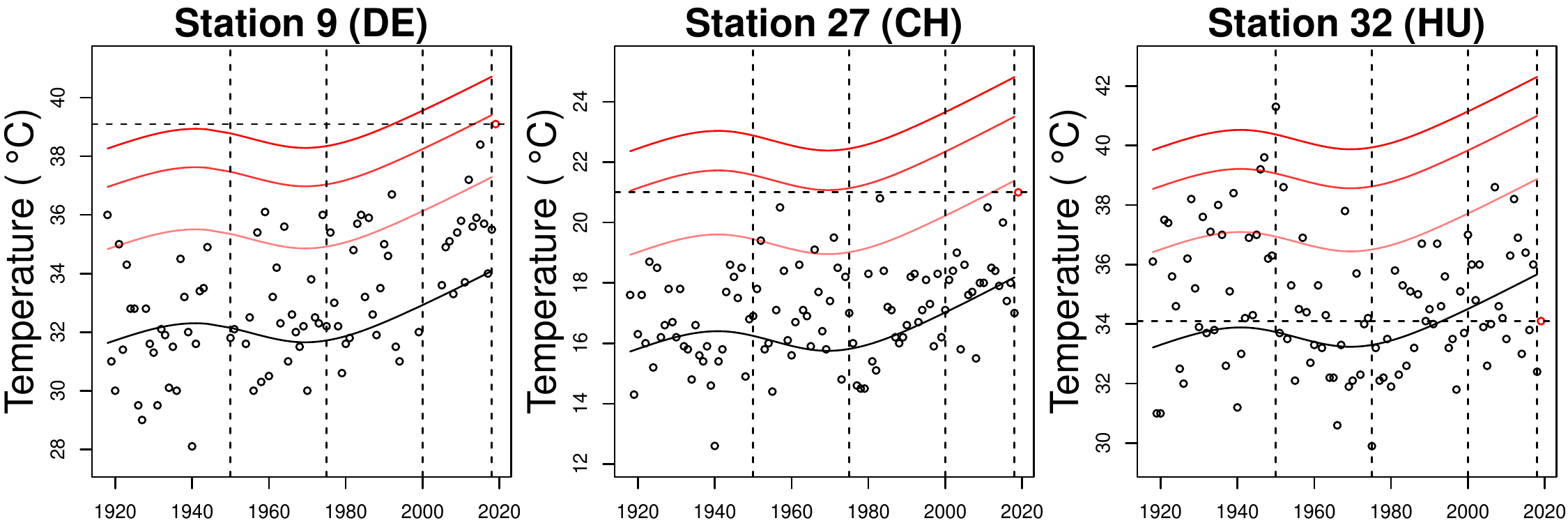}
\caption{Plot of observed annual maxima (dots), estimated time trend (solid black), and estimated 10, 100 and 1000 year-return level curves (from light to dark red) for 3 selected stations located in Germany, Switzerland and Hungary (left to right). The red dot in each panel represents the observed annual maximum for 2019 (not used to fit the model). The four vertical dashed lines correspond to the reference years of 1950, 1975, 2000, and 2018.}
\label{fig:return_level}
\end{figure}

We then examine the fitted time trend, as well as the estimated time-varying $M$-year return level $z^M_{\bm s,t}$, defined for each site $\bm s\in\calS$ as the $(1-1/M)$-quantile from the fitted GEV distribution, i.e., $z^M_{\bm s,t}=\hat\mu_{\bm s,t}-\hat\sigma[\{-\log(1-1/M)\}^{-\hat\xi}-1]/\hat\xi$. Under temporal stationary conditions, the $M$-year return level is expected to be exceeded once every $M$ years (at each site). With global warming, return levels from the past may be exceeded much more frequently in the present and future. In other words, observations that were extreme in the past may no longer be as rare under the current conditions. The effect of climate change can thus be assessed based on return levels. Figure~\ref{fig:return_level} exhibits the estimated time trend and the corresponding 10, 100 and 1000 year-return level curves for 3 selected stations located in Germany, Switzerland and Hungary. The estimated time trend is at its lowest around 1975 and its highest in 2018, which corroborates other studies about climate change. The red dot in each plot represents the observed annual maximum for 2019 (not used to fit the model). For station 9, the 2019 annual maximum exceeds the 1000 year-return levels corresponding to 1950 and 1975. However, it barely reaches the 100 year-return level for 2018. For station 27, the 2019 annual maximum approximately corresponds to the 100-year event when taking 1975 as the reference year, but it becomes a 10-year event when taking 2018 as the reference. For these two stations in Germany and Switzerland, the 2019 heatwave was therefore very extreme compared to mid-20th century conditions, but only moderately extreme with respect to current climate. As for the station 32 in Hungary, our model suggests that the 2019 annual maximum was not very extreme overall (both with respect to past and current conditions). 


\subsection{Spatial dependence structure and model comparison}
We now use the estimated marginal distributions and transform the data to the standard uniform ${\rm Unif}(0,1)$ scale. We next estimate the dependence structure (i.e., the copula) through maximum pairwise likelihood inference using the max-id model introduced in \S\ref{sec:newmodel}--\ref{sec:nonstatcorrel}. The most complex model that we fit is the non-stationary, but locally isotropic dependence structure specified in \S\ref{sec:nonstatcorrel} and Table~\ref{tab:sub-models}, and we also fit several sub-models for comparison. Specifically, our most general model assumes that $\Omega_{\bm s,t}(R_i)$ in \eqref{eq:Omega} has the form 
$\Omega_{\bm s,t}(R_i) = \lambda_{\bm s,t}^2(1+R_i)^{-2\nu}I_{2\times2}$, $\lambda_{\bm s,t} = \exp(\lambda_0+\lambda_1\times\text{alt}_{\bm s}+\lambda_2\times t)$. 
Using \eqref{eq:maxidconstr} for the mean measure of the Poisson points $\{R_i\}$ arising in the spectral representation \eqref{max-stable}, the parameter vector to be estimated is thus 
$\bm\psi=(\alpha,\beta,\lambda_0,\lambda_1,\lambda_2,\nu)^T\in(0,\infty)^2\times\Real^4$. 
We compare this model to the fits of five simpler models, contained as special cases (or limiting cases) of our model, leading to features such as stationarity or max-stability. Specifically, Model 1 corresponds to the stationary extremal-t max-stable process, and Model 3 to a stationary max-id model proposed by \citet{Huser.etal:2020}. Models 2 and 4 are their non-stationary counterparts. Finally, Models 5 and 6 are our new stationary and non-stationary max-id models, with an explicit magnitude-dependent range of dependence. These six different models are specified with the following parameter configurations:
$$
\left\{
\begin{array}{lll}
\text{Model 1}:& \{\alpha>0,\beta\downarrow0,\lambda_0\in\Real,\lambda_1=0,\lambda_2=0,\nu=0\},&\text{stationary max-stable}\\
\text{Model 2}:& \{\alpha>0,\beta\downarrow0,\lambda_0\in\Real,\lambda_1\in\Real,\lambda_2\in\Real,\nu=0\},&\text{non-stationary max-stable}\\
\text{Model 3}:& \{\alpha>0,\beta>0,\lambda_0\in\Real,\lambda_1=0,\lambda_2=0,\nu=0\},&\text{stationary simple max-id}\\
\text{Model 4}:& \{\alpha>0,\beta>0,\lambda_0\in\Real,\lambda_1\in\Real,\lambda_2\in\Real,\nu=0\},&\text{non-stationary simple max-id} \\
\text{Model 5}:& \{\alpha>0,\beta>0,\lambda_0\in\Real,\lambda_1=0,\lambda_2=0,\nu\in\Real\},&\text{stationary general max-id}\\
\text{Model 6}:& \{\alpha>0,\beta>0,\lambda_0\in\Real,\lambda_1\in\Real,\lambda_2\in\Real,\nu\in\Real\},&\text{non-stationary general max-id}  
\end{array}\right.
$$
To assess the uncertainty of estimated parameters, we used the parametric bootstrap procedure with $300$ bootstrap samples for each model as described in \S\ref{sec:PL}. The  estimates and the $95\%$ bootstrap confidence intervals are reported in Table~\ref{CI}. The estimates for $\alpha$ and $\beta$ are relatively large with lower confidence bounds clearly above $0$, indicating that the data are asymptotically independent.
Moreover, in Models 5 and 6, we obtain relatively large estimates $\hat\nu$ with lower confidence bounds above $1$ and $2$, respectively, which suggests that the range of spatial dependence is substantially smaller for more severe extreme events. In all non-stationary models, the estimates for the altitude coefficient $\lambda_1$ are significantly negative, such that the range of dependence diminishes in subregions with higher altitudes. From our new Model 6, $\hat\lambda_1=-0.31$, so the spatial extent of heatwaves is estimated to be about $\exp(0.31)\approx1.36$ smaller 1km higher (in altitude). The estimates of $\lambda_2$ are positive in all three non-stationary models, hinting that the spatial extent of heatwaves has increased in recent years, and Model 6 suggests that it increases by a factor about $\exp(0.23)\approx1.26$ per century. However, this effect is not significant based on the available data.
\begin{table}[t!]
\footnotesize
\centering
\caption{Parameter estimates for the six max-id models fitted to annual European temperature maxima, with $95\%$ confidence intervals (indicated as subscripts) based on the parametric bootstrap procedure described in \S\ref{sec:PL} using $300$ replications. Here, $\hat\lambda_1$ and $\hat\lambda_2$ represent the increase in $\log\hat\lambda_{\bm s,t}$ (log-range) per km in altitude, and per century in time, respectively.}\label{CI}
\vspace{5pt}
\begin{tabular}{c|c|c|c|c|c|c}
   & $\hat\alpha$  & $\hat\beta$ & $\hat\lambda_0$ & $\hat\lambda_1$ & $\hat\lambda_2$ & $\hat\nu$ \\
\hline
Model 1& $5.0_{(3.5, 10.0)}$ & 0 & $0.04_{(-0.31, 0.71)}$ & 0 & 0 & 0 \\
Model 2& $5.1_{(3.7, 10.0)}$ & 0 & $0.09_{(-0.29, 0.92)}$ & $-0.31_{(-0.44, -0.13)}$ & $0.31_{(-0.40, 0.89)}$ & 0\\
Model 3& $2.5_{(0.5, 6.5)}$ & $1.5_{(0.4, 3.9)}$ & $-0.35_{(-0.60, 0.19)}$ & 0 & 0 & 0\\
Model 4& $2.5_{(0.6, 6.4)}$ & $1.5_{(0.3, 4.2)}$ & $-0.28_{(-0.55, 0.45)}$ & $-0.40_{(-0.56, -0.17)}$ & $0.30_{(-0.43 ,0.76)}$ & 0\\
Model 5& $5.0_{(0.5, 9.9)}$ & $2.3_{(1.1, 9.8)}$ & $1.85_{(0.60, 3.88)}$ & 0 & 0 & $2.9_{(1.2, 6.0)}$\\
Model 6& $5.5_{(2.7, 8.3)}$ & $2.4_{(1.0, 7.3)}$ & $2.12_{(1.71, 2.92)}$ & $-0.31_{(-0.43, -0.12)}$ & $0.23_{(-0.55, 0.83)}$ & $3.2_{(2.6, 4.3)}$ \\
\end{tabular} 
\end{table}

To assess the relative goodness-of-fit and test the predictive performance of the six models, we use a cross-validation scheme, whereby each station $\bm s_{j_0}$, $j_0=1,\dots,D$, is left out at a time and the six models refitted. We then compare the models using the logarithmic score, 
\begin{equation}\label{eq:logS}
        {\rm LogS}_{j_0}=\sum_{j\neq j_0}\left[\sum_{k=1}^n V(z_{kj},z_{kj_0})-\log\{V_1(z_{kj},z_{kj_0})V_2(z_{kj},z_{kj_0})-V_{12}(z_{kj},z_{kj_0})\}\right],
\end{equation}
which is the sum of the log pairwise-densities by considering only the pairs composed of the left-out station $\boldsymbol{s}_{j_0}$ and one of the other stations $\boldsymbol{s}_j$, $j\neq j_0$. Logarithmic scores are strictly proper in the sense of \citet{Gneiting2012},  such that they enable us to appropriately compare the predictive power of different models. The final score of a model is obtained by summing scores for all stations, i.e., using ${\rm LogS} = \sum_{j_0=1}^D{\rm LogS}_{j_0}$. 
In our model comparison, we also include traditional geostatistical models from the spatial statistics literature, which do not have the strong theoretical motivation from Extreme-Value Theory. Precisely, we also fit the Gaussian copula and the Student-$t$ copula models with $\alpha>0$ degrees of freedom, using the same stationary or non-stationary correlation function as before. For consistency, we use the same pairwise likelihood inference approach. We label these models as follows: Model 7 is the stationary Gaussian copula model; Model 8 is its non-stationary counterpart; Model 9 is the stationary Student-$t$ copula model; Model 10 is its non-stationary counterpart. 

\begin{table}[t!]
\centering
\caption{Ranking of the 10 models using the cross-validated logarithmic score \eqref{eq:logS} for pairwise predictions. Lower rank means better predictive performance.}\label{tab:table2}
\vspace{5pt}
\resizebox{\textwidth}{!}{
\begin{tabular}{cccccccccc}
	\hline
	Model 1& Model 2& Model 3& Model 4& Model 5& Model 6& Model 7& Model 8& Model 9& Model 10 \\
	9&8&4&3&2&1&10&7&6&5 \\
	\hline
\end{tabular}%
}
\end{table}

The final ranking of all models based on the logarithmic score is reported in Table~\ref{tab:table2}. Interestingly, the ``traditional" models from spatial statistics and Extreme-Value Theory, namely the Gaussian copula (Models 7, 8) and max-stable (Models 1, 2) models, perform worst. Furthermore, the non-stationary Gaussian copula (Model 8) outperforms its max-stable counterpart (Model 2) despite the additional parameters of the latter, which casts strong doubts about the max-stability assumption and suggests that the dependence strength of maxima weakens at higher quantiles. The four estimated max-id (but not max-stable) models have the best results, and the most complex model that we propose (Model 6), which includes covariate effects of altitude and time, as well as the magnitude-dependent probabilistic structure, performs the best overall. Finally, the non-stationary Student-$t$ copula (Model 10) ranks 5th, right behind the max-id (non-max-stable) models. Its flexible structure---being at the same time in the domain of attraction of the max-stable extremal-$t$ limit, and also very close to the Gaussian copula for large degrees of freedom---seems to compensate for some of the weaknesses of max-stable and Gaussian copula models. 

\begin{figure}[t!]
    \centering
    \includegraphics[width=\textwidth]{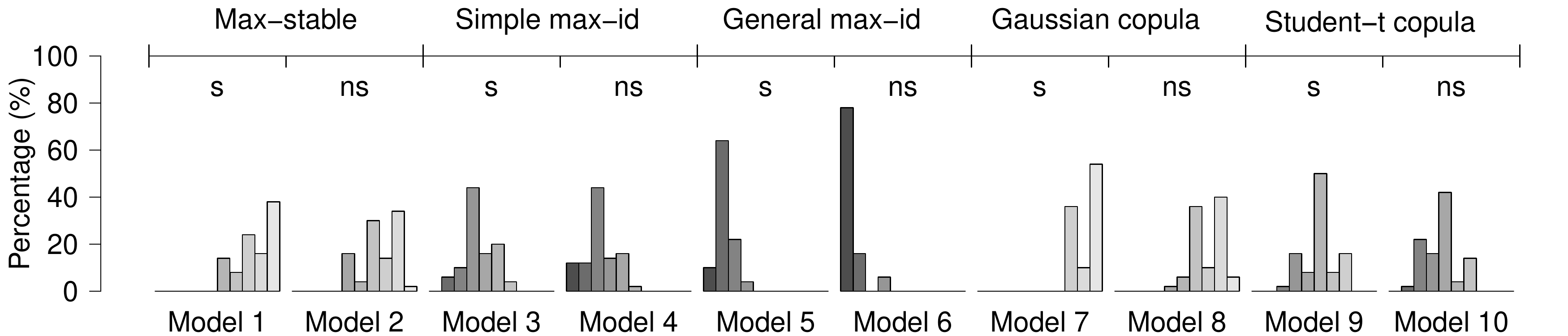}
    \caption{Logarithmic score \eqref{eq:logS} ranks computed for $50$ bootstrap simulations from the fitted Model 6, for each of the 10 models. The bars show the percentage of ranks ranging from 1 (best score) to 10 (worst score) for each of the Models 1--10, with darker grey (towards the left side of each histogram) indicating better rank. The letter ``s'' means ``stationary'', whereas ``ns'' means ``non-stationary''.}
    \label{fig:compare_models}
\end{figure}

We then conduct a bootstrap simulation experiment, in order to confirm our conclusions from this model comparison, assess the uncertainty of the ranking,  and remove any model selection bias. Precisely, we simulate 50 datasets according to the best model (Model 6), where we use the same sample size and structure of missing values as in the real dataset. For each of the $50$ simulated datasets, we then refit the 10 different models and recompute the ranking based on the logarithmic score, ${\rm LogS}$. This gives $50$ rankings for the models $1$--$10$. Figure~\ref{fig:compare_models} shows the percentage of times that a given model was ranked $1$st to $10$th. Models with high bars towards the left are generally better. We clearly see that our most complex max-id models (Models 5 and 6) have the best performances, and Model 6 is ranked $1$st overall in about 40 out of the 50 cases. This Monte Carlo experiment therefore confirms our initial findings and the advantage of the very flexible dependence structure of our proposed Model 6 with respect to the other models. 

\begin{figure}[t!]
    \centering
    \includegraphics[width=\textwidth]{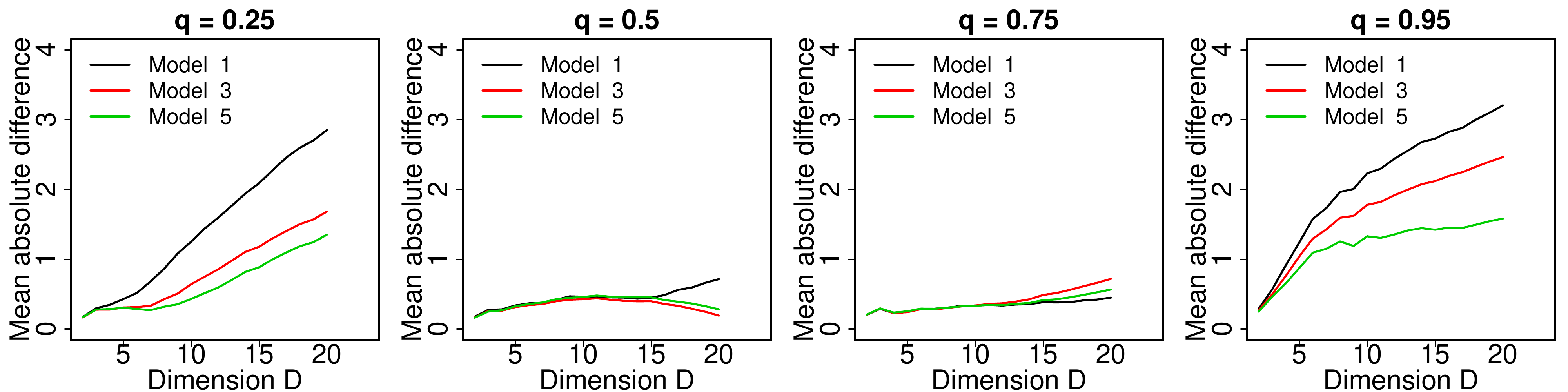}
    \caption{Mean absolute difference between empirical and fitted extremal coefficients $\theta_D(z)$, plotted with respect to dimension $D=2$--$20$, for Models 1 (black), 3 (red), and 5 (green), at unit Fr\'echet quantile levels $z=-1/\log(q)$ with $q=0.25, 0.5, 0.75$ and $0.95$ (left to right).}
    \label{fig:mean_diff}
\end{figure}

If a model appropriately captures the dependence structure of the data, it is expected that the fitted extremal coefficients $\hat\theta_D(z)$ from the model are close to the empirical extremal coefficients $\hat\theta_D^{\text{emp}}(z)=-z\log[\hat\Pr\{Z(\bm s_1)\leq z,\dots,Z(\bm s_D)\leq z\}]$ at level $z$ (assuming here unit Fr\'echet marginals), where $\hat\Pr$ is the empirical probability. Since Models 2, 4 and 6 are non-stationary, empirical extremal coefficients are more tricky to estimate accurately in these cases. Therefore, for simplicity, we here only compare the fitted extremal coefficients of Models 1, 3 and 5, which are the stationary versions of the max-stable model, the simple max-id model of \citet{Huser.etal:2020} and our proposed general max-id model, respectively, with their empirical counterparts in dimensions $D=2$--$20$. In dimensions $D=2$ and $3$, we computed extremal coefficients for all pairs and triplets of the 44 stations, whereas in higher dimensions, we only computed coefficients for a maximum of 1000 randomly sampled combinations of stations among the 22 stations without missing values. Figure~\ref{fig:mean_diff} shows the average absolute difference between the empirical and fitted extremal coefficients $\theta_D(z)$ in dimensions $D=2$--$20$ at unit Fr\'echet quantile levels $z=-1/\log(q)$ with $q=0.25, 0.5, 0.75$ and $0.95$ for Models 1, 3 and 5. Notice that under stationarity, these levels are on average marginally exceeded 3 times in 4 years, once in 2 years, once in 4 years, and once in 20 years, respectively, so they correspond to moderately extreme events.
All three models are comparable for moderate quantiles $q=0.5$ and $q=0.75$ representing the behavior in the bulk of the max-id distribution. The relatively complex max-id Model 5 (green curve) performs sensibly better than the max-stable model (black curve) and the simple max-id model (red curve) at quantile levels $q=0.25$ and $0.95$, especially in higher dimensions. Model 5 thus better captures the dependence structure of spatial extreme events of relatively small and large magnitudes. Throughout, the observed absolute differences are not excessively large compared to the theoretical range $[1,D]$ of extremal coefficients.

\begin{figure}[t!]
    \centering
    \includegraphics[width=\textwidth]{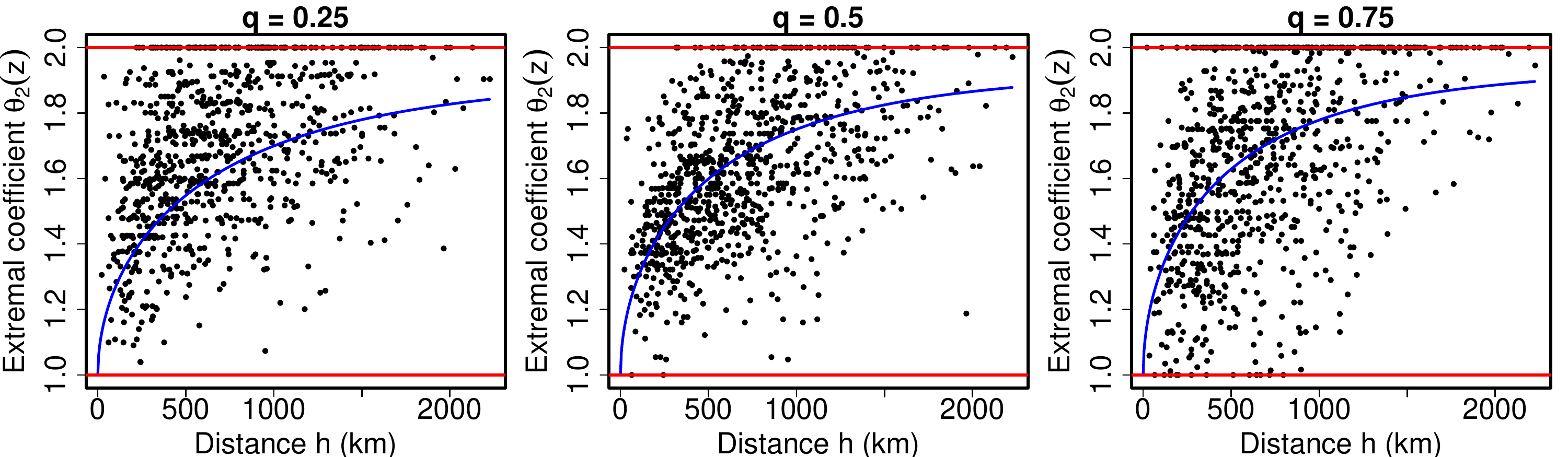}
    \caption{Bivariate empirical extremal coefficients $\theta_2(z)$ for all pairs of sites (black), plotted with respect to spatial distance, and theoretical curve (blue) based on the fitted Model 5, for unit Fr\'echet quantile levels $z=-1/\log(q)$ with $q=0.25, 0.5$ and $0.75$ (left to right).}
   \label{fig:ext_model_2b}
\end{figure}

\begin{figure}[t!]
    \centering
    \includegraphics[width=\textwidth]{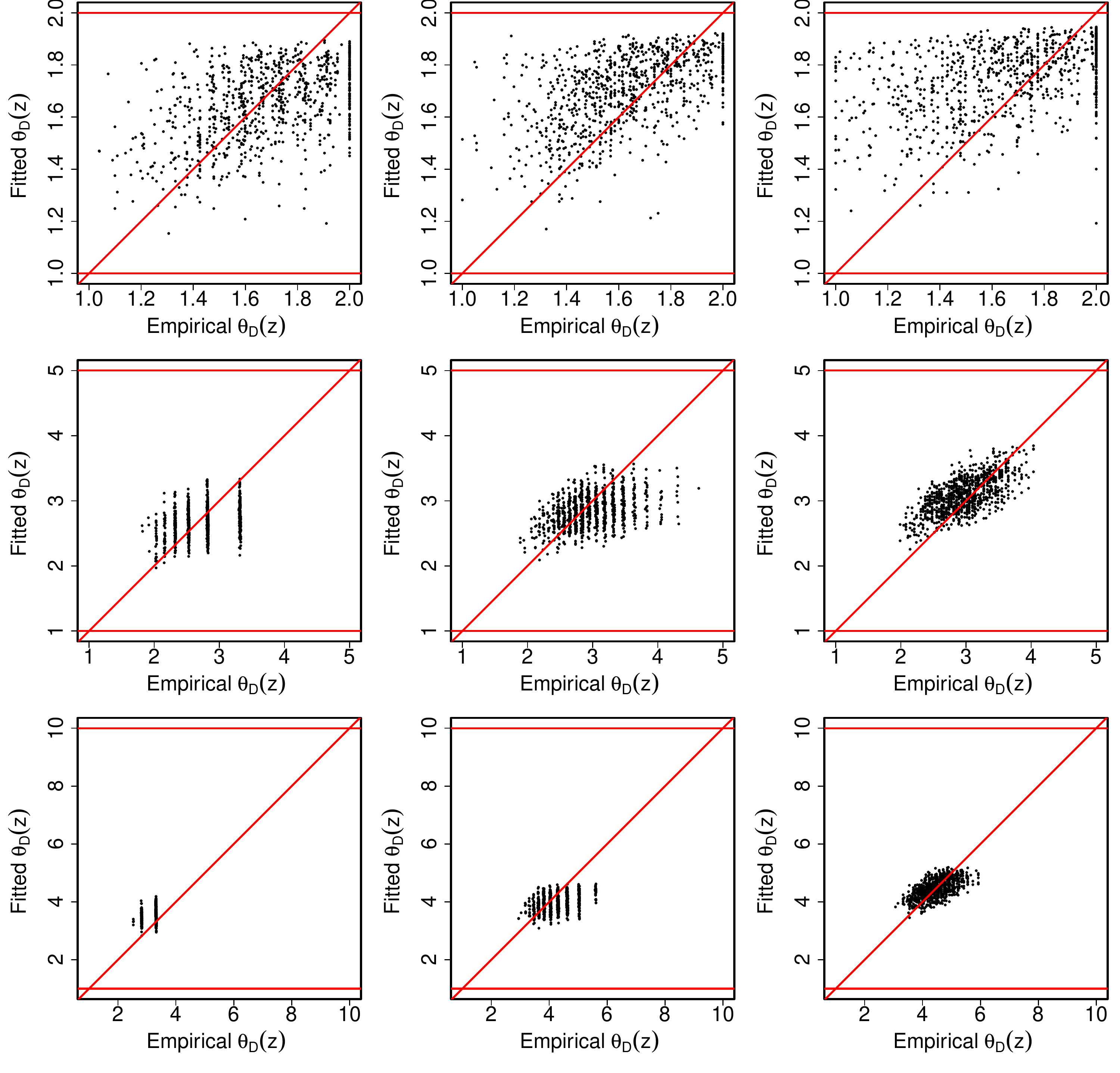}
    \caption{Scatterplots of empirical versus fitted extremal coefficients $\theta_D(z)$ for Model 5 in dimensions $D=2, 5, 10$ (top to bottom) for unit Fr\'echet quantile levels $z=-1/\log(q)$ with $q=0.25, 0.5,0.75$ (left to right). The main diagonal indicates a perfect fit.}
   \label{fig:ext_model_2a}
\end{figure}

To further assess the goodness-of-fit and verify the fidelity of our fitted max-id Model 5 to the data, Figure~\ref{fig:ext_model_2b} compares bivariate empirical coefficients $\theta_2(z)$, plotted with respect to spatial distance, to their model-based counterparts, for three different quantile levels $z$. 
Although the variability of bivariate empirical extremal coefficients is high, the fitted curves seem to adequately capture the decay of spatial dependence with distance. Our fitted model suggests that extremal dependence persists at very large distances, which is consistent with heatwaves being large-scale phenomena with the potential of simultaneously affecting large parts of Europe. We also verify the goodness-of-fit in higher dimensions. Figure~\ref{fig:ext_model_2a} shows scatterplots of empirical versus fitted extremal coefficients $\theta_D(z)$ for Model 5 in dimensions $D=2,5,10$ for unit Fr\'echet quantile levels $z=-1/\log(q)$ with $q=0.25, 0.5$ and $0.75$. 
The dots tend to concentrate around the main diagonal, especially in high dimensions, which confirms a satisfactory model fit. Nevertheless, the fitted model tends to be slightly smoother in general than empirical data in terms of the range of values of empirical coefficients, but such behavior can be expected since our model cannot perfectly capture all the non-stationarities of extremal dependence arising over this very large and geographically heterogeneous study region. While the stationary max-id Model 5 already produces a very decent fit, our non-stationary Model 6 is expected to perform even better.

\begin{figure}[t!]
    \centering
    \includegraphics[width=\textwidth]{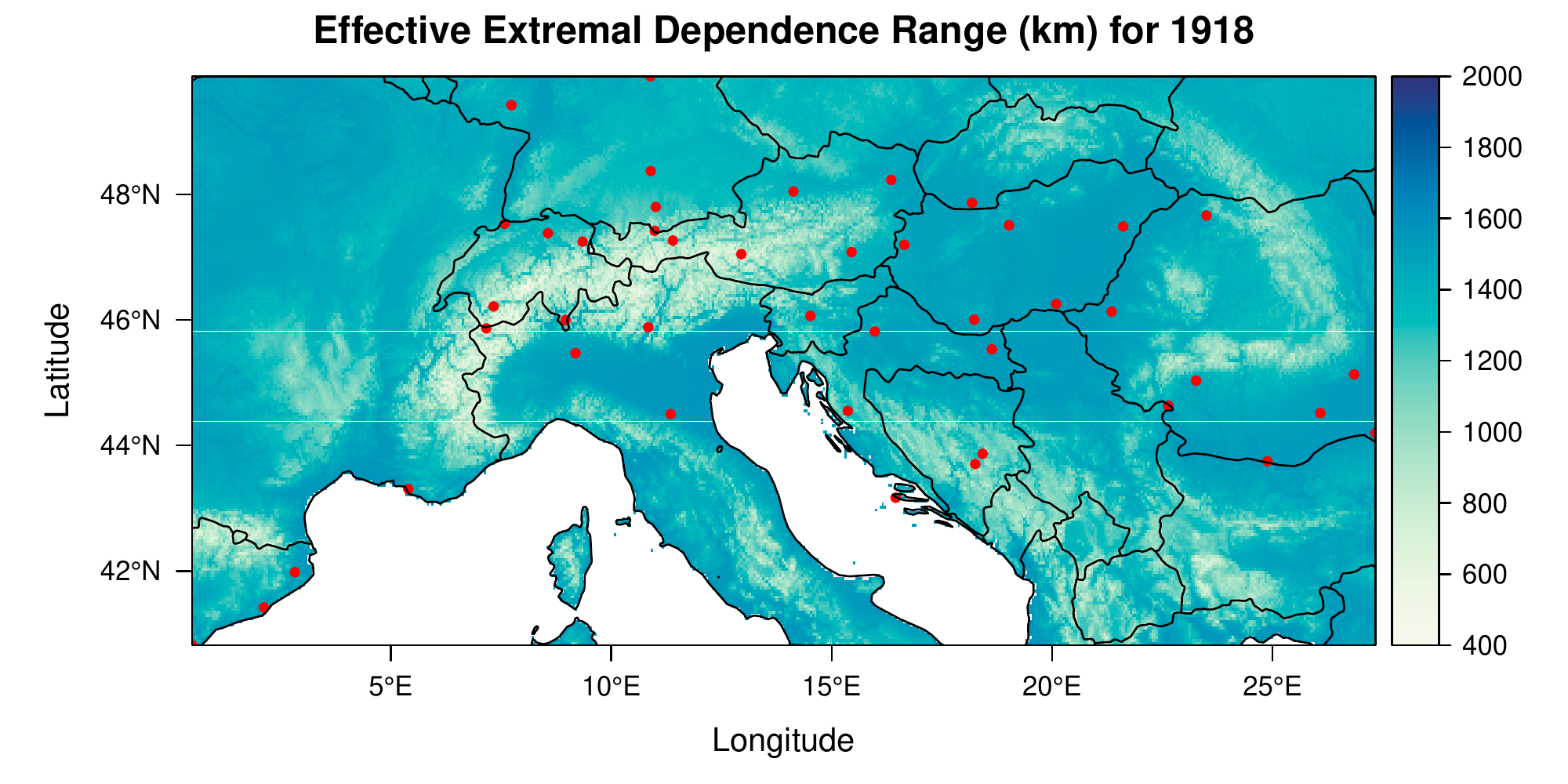}
    \includegraphics[width=\textwidth]{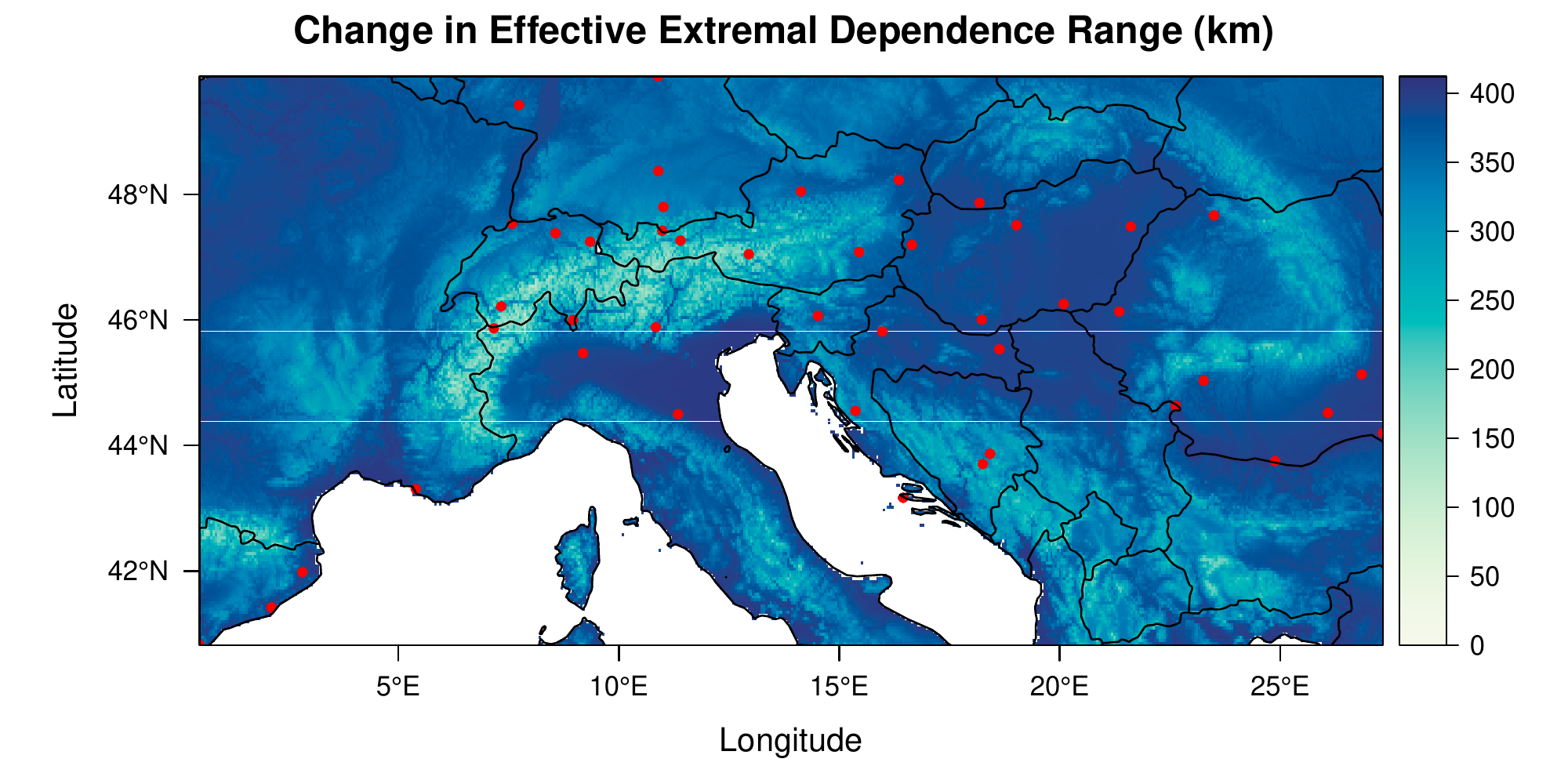}
    \caption{Map of the effective extremal dependence range (km) for 1918 (top), and of the difference between 2018 and 1918 (bottom), at the level $z=-1/\log(0.9)$. Monitoring stations are shown as red dots.}
   \label{fig:maprange}
\end{figure}

In order to visualize the spatio-temporal variation in the estimated extremal dependence structure, and to assess whether the spatial extent of heatwaves has changed a lot over time due to climate change, we then compute the effective extremal dependence range for 1918 and 2018, based on the fitted non-stationary Model 6. We define the effective extremal dependence range (at a given point in space and time) as the minimum spatial distance (from that point) such that $\theta_2(z)=1.95$ for a given level $z$, under constant covariate values. Figure~\ref{fig:maprange} displays a map of the results for 1918 taking $z$ as the level $z=-1/\log(0.9)$, as well as the difference between the results for 2018 and 1918.
From the top panel, we can see that the effective extremal dependence range varies from about $400$km at high altitudes to $1500$km at low altitudes. Altitude is thus a major (significant) covariate. From the bottom panel, we see that our Model 6 estimates the change in extremal dependence range over the last century to be between about $150$km at high altitudes and $400$km at low altitudes. Heatwaves might therefore have become slightly larger in extent, especially at low altitudes.

\subsection{Probabilistic assessment of the 2019 European heatwaves}\label{sec:2019heatwave}
\begin{figure}
\centering
\includegraphics[width=\textwidth]{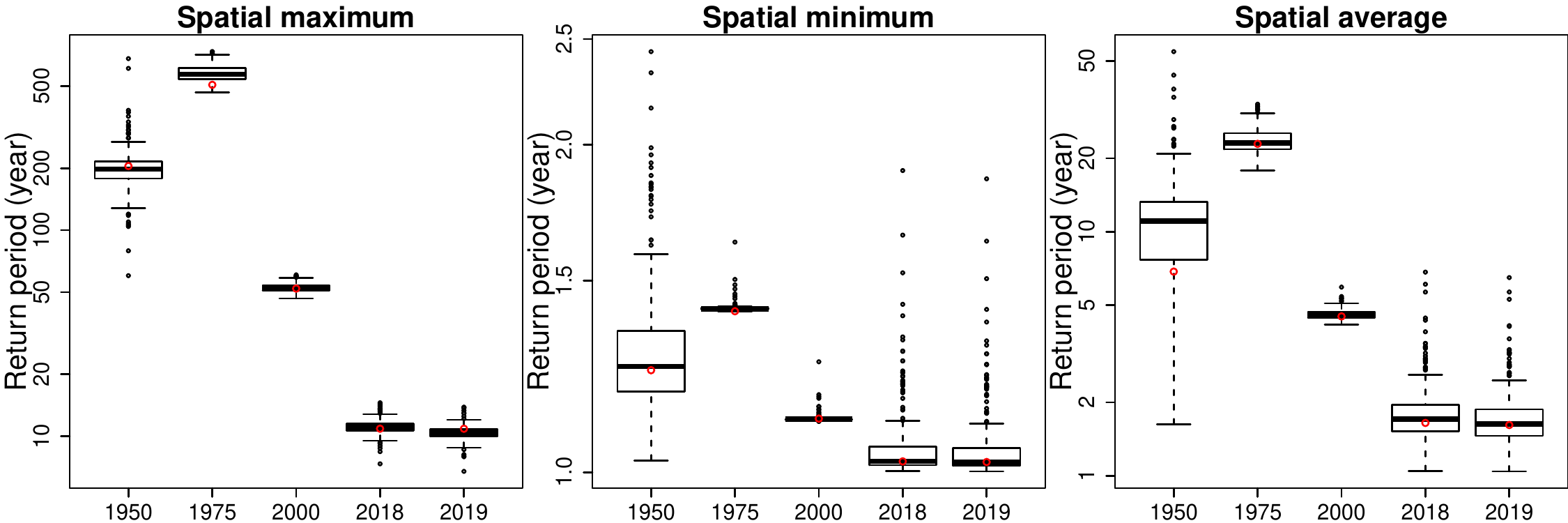}
\caption{Boxplots of bootstrapped return periods (on logarithmic scale) of the 2019 annual maximum, computed for the spatial maximum, minimum and average (left to right) over 31 stations, based on $10^5$ random fields simulated from our best non-stationary spatial Model~6 (fitted to the annual maxima for the period $1918$--$2018$), with respect to 1950, 1975, 2000, 2018 and 2019 as reference years. Red dots are pointwise estimates of these return periods.}
\label{fig:boxplot_2019}
\end{figure}

We conclude our real data analysis with a probabilistic assessment of the extremes observed during the 2019 European heatwaves, which affected large parts of Europe. Over the summer 2019, many monitoring stations across Europe indeed recorded the highest temperature in almost a century. A natural question is whether this could have been anticipated from historical data. To assess the severity of the 2019 European heatwaves, we here complement the marginal analysis of \S\ref{sec:spatialtrends}, by simulating $10^5$ replicates from our best fitted non-stationary Model 6 at 31 stations for which the 2019 annual maxima are available, and transforming these simulated data to their estimated marginal GEV scales. From these $10^5$ replicates, we then compute empirical return periods for the spatial maximum, spatial minimum and spatial average of the observed 2019 maxima, with respect to the reference years $1950$, $1975$, $2000$, $2018$ and $2019$. To estimate the variability of our return period estimates, we use the 300 bootstrap fits and recompute these return periods. Figure~\ref{fig:boxplot_2019} shows boxplots of the bootstrapped return periods, as well as the point estimates (red dots). Due to the estimated time trend (both in margins and dependence), return periods are always highest when compared to 1975 and lowest when compared to 2018--2019. When considering return periods for the spatial maximum (left panel), which is large when at least one site experiences an extreme event, we get a return period of about 500 years when compared to the climatic conditions of 1975, but only about 10 years when compared to current climatic conditions. When considering the spatial average, we get a return period of about 20--30 years when compared to 1975, but only 2 years for 2018--2019. Finally, when considering the spatial minimum, which is large only when all sites experience simultaneous extreme events, and which is  usually observed at one of the locations in the Alps,  the 2019 heatwaves were not especially extreme, corresponding only to a 1--1.5 year event for all reference years. 


\section{Conclusion}
\label{sec:conclusion}

We have proposed a non-stationary max-id spatial model for block maxima, which embeds spatio-temporal covariates in its dependence structure, while having a very flexible form of weakening dependence strength at increasingly high quantiles, in order to model extreme temperatures over Southern Europe. Our fitted models reveal that the dependence structure of temperature annual maxima is significantly weaker at higher altitudes, and similarly for more severe heatwaves. The estimated parameters of our models with temporal non-stationary also suggest that the spatial extent of heatwaves has become wider in recent years, though this effect was not significant based on our parametric bootstrap procedure. 

Modeling approaches in classical Gaussian-based geostatistics and spatial extreme-value analysis often use a setting where the dependence structure is stationary over both space and time. This assumption is problematic when spatial and temporal scales are large and lead to heterogeneous regional and temporal characteristics in co-occurrence patterns of extreme values, and even more so when we aim to detect and analyze such patterns. The max-id models developed in this paper are a step forward towards more accurate inference while keeping parsimonious specifications. Trends in dependence are notoriously difficult to estimate when data are not abundant, and one has to carefully avoid confusion with marginal trends. Indeed, the accurate modeling of marginal trends in extremes remains of paramount importance, and it is a prerequisite to avoid estimating spurious trends in dependence models. In our real data application, we implemented semi-parametric spline functions for capturing marginal trends in the GEV parameters, and we opted for a flexible tensor product specification to allow interaction of trends arising in latitude, longitude and altitude. We also assessed spatial return periods associated with the 2019 Europe heatwaves over Southern Europe, and concluded that the summer 2019 was very extreme when considering the spatial maximum over the monitoring stations (especially compared to mid-20th century conditions), moderately extreme when considering the spatial average, and not especially extreme when considering the spatial minimum. Furthermore, our analysis provided clear evidence for climate change and its impact on spatial extreme temperature events.

Finally, we underline the main methodological novelty of building magnitude-dependent max-id models, where the spatial dependence range becomes shorter as events become more extreme. Our construction \emph{explicitly} accounts for this behavior, and allows us to capture in a single parsimonious parametric model: (i) max-stable asymptotic dependence; (ii) weakening asymptotic dependence; (iii) weakening asymptotic independence. By keeping a flexible max-stable process on the boundary of the parameter space, our proposed model achieves the subtle trade-off of combining the strength of theoretically-motivated max-stable models together with the pragmatism of flexible max-id extensions with weakening dependence strength. Our sophisticated extreme-value model, combined with covariates and geometric anisotropy, thus provides a very rich class of models for spatially-indexed block maxima, and opens the door to more realistic risk assessment of extreme environmental events. 

\baselineskip=14pt

\bibliographystyle{CUP}
\bibliography{reference}

\baselineskip 10pt

\end{document}